\documentclass[aip,jcp,reprint]{revtex4-1}

\usepackage{amsmath}
\usepackage{amsfonts}
\usepackage{amssymb}
\usepackage{graphicx}
\usepackage{moreverb}
\usepackage{url}
\usepackage{epsf}
\usepackage{color}
\usepackage{multirow}
\usepackage{subfigure}

\newcommand{\rr}{{\bf r}}

\newcommand{\gn}{\nabla n}
\newcommand{\DZPmv}{$^{(P)}\mathrm{d}\zeta+\mathrm{p}$}
\newcommand{\QZDP}{$\mathrm{q}\zeta+\mathrm{dp}$}

\begin{document}

\title{Room temperature compressibility and diffusivity of liquid water from first principles}
\author{Fabiano~Corsetti}
\email[E-mail: ]{f.corsetti@nanogune.eu}
\affiliation{CIC nanoGUNE, 20018 Donostia-San Sebasti\'{a}n, Spain}
\author{Emilio~Artacho}
\affiliation{CIC nanoGUNE, 20018 Donostia-San Sebasti\'{a}n, Spain}
\affiliation{Theory of Condensed Matter, Cavendish Laboratory, University of Cambridge, Cambridge CB3 0HE, United Kingdom}
\affiliation{Basque Foundation for Science Ikerbasque, 48011 Bilbao, Spain}
\affiliation{Donostia International Physics Center, 20018 Donostia-San Sebasti\'{a}n, Spain}
\author{Jos\'{e} M.~Soler}
\affiliation{Dept. de F\'{i}sica de la Materia Condensada, Universidad Aut\'{o}noma de Madrid, 28049 Madrid, Spain}
\author{S.~S.~Alexandre}
\affiliation{Dept. de F\'{i}sica, Universidade Federal de Minas Gerais, 30123-970 Belo Horizonte, Brazil}
\author{M.-V.~Fern\'{a}ndez-Serra}
\affiliation{Dept. of Physics and Astronomy, Stony Brook University, Stony Brook, New York 11794-3800, USA}
\date{\today}

\begin{abstract}
The isothermal compressibility of water is essential to understand its anomalous properties. We compute it by {\em ab initio} molecular dynamics simulations of 200 molecules at five densities, using two different van der Waals density functionals. While both functionals predict compressibilities within $\sim$30\% of experiment, only one of them accurately reproduces, within the uncertainty of the simulation, the density dependence of the self-diffusion coefficient in the anomalous region. The discrepancies between the two functionals are explained in terms of the low- and high-density structures of the liquid.
\end{abstract}

\maketitle

\section{Introduction}

To date, most of the experimentally measured anomalies of water have been reproduced in molecular dynamics or Monte Carlo simulations using empirical force fields~\cite{Guillot2002}, albeit with significant differences in the predictions given by different models~\cite{Paschek2004,Abascal2005,Vega2009}. These simulations are the main contributors to the debate~\cite{Sciortino1997,Moore2011,Chandler2012,MallamacePNAS2013,Huang2009} on the existence of a liquid-liquid critical point (LLCP), postulated to explain its anomalous response functions, which show a divergent behavior in the supercooled phase. Many studies of these response functions, both in the supercooled~\cite{MallamacePNAS2013,Kumar2011,Sciortino2011,Abascal2011} and high temperature~\cite{Paschek2004,Vega2009} regions of the liquid, have been published in the last five years. Although many simulations find this second critical point at high $P$ and low $T$~\cite{Paschek2005,Gallo2010,Abascal2010,Sciortino2011}, it is still open whether the LLCP is a simulation-dependent feature, and how accurately these empirical force fields capture the correct physics of the hydrogen bonds~\cite{Pamuk2012}.

In principle, {\em ab initio} molecular dynamics (AIMD), based on density-functional theory (DFT), could be used to validate certain structural and dynamical properties of these models. In practice, they have not yet been able to contribute much to the discussion. In fact, simulations using standard exchange and correlation (xc) semi-local (GGA) functionals were not even able to reproduce the structure and diffusivity of water at room temperature~\cite{Grossman2004,Fernandez-Serra2004,Kuo2004,Sit2005,Schmidt2009}. The development of new functionals~\cite{vdW-DF,Klimes2010,Lee2010,VV10} that account for van der Waals (vdW) interactions from first principles is changing this trend, with promising results for both liquid water~\cite{Lin2009,Mogelhoj2011,water_emiliomarivi,Zhang2011a,Wikfeldt2011} and ice~\cite{Pamuk2012,Murray2012}.

Beyond the LLCP discussion, an accurate first principles description of liquid water is needed to simulate heterogeneous systems such as the metal/water interface, or the water/semiconductor interface, relevant for electro- and photo-catalytic applications. In both cases, an accurate and explicit quantum-mechanical description of the chemistry at the interface needs to be accounted for. However, questions such as how much the equilibrium density of the simulated water affects the interfacial electronic and atomic structure of the simulated systems have not yet been explored, because very little is known about the phase diagram of liquid water using different xc functionals.

In this paper, we present an extensive series of AIMD simulations of cells of up to 200 molecules of liquid water using two non-local vdW density functionals: the vdW-DF functional of Dion {\em et al.}~\cite{vdW-DF}, and the VV10 form of Vydrov and Van Voorhis~\cite{VV10}. From our large-scale simulations we extract smooth pressure--density ($P$--$\rho$) equations of state, finding compressibilities within $\sim$30\% of experimental measurements. Even more importantly, one of the functionals accurately predicts the maximum of diffusivity as a function of density. This represents an important validation of vdW-DF-based AIMD simulations. Furthermore, the dynamics of the H-bond network near this anomaly can be used to analyze and evaluate classical force fields.

\section{Computational methods}

\subsection{AIMD simulations}

We employ the SIESTA~\cite{Soler2002} code, with norm-conserving pseudopotentials in Troullier-Martins form~\cite{Troullier1991} and a basis set of numerical atomic orbitals (NAOs) of finite support. We employ a variationally-obtained~\cite{Junquera2001,Anglada2002} double-$\zeta$ polarized basis (which we refer to as \DZPmv\ for consistency with Ref.~[\onlinecite{Corsetti2013}]). Our AIMD simulations use a time step of 0.5 fs and (unless otherwise stated) 200 molecules of heavy water. However, the reported mass densities are rescaled to those of light water for ease of comparison. The initial geometry is obtained from a classical MD run of 1~ns, using the TIP4P force field~\cite{Jorgensen1983b} in the GROMACS~\cite{gromacs} code, followed by an AIMD equilibration run of 3~ps, using velocity rescaling at 300~K, and a production run of 20~ps, using constant-energy Verlet integration.

Additionally, we perform a number of smaller simulations, of 64 and 128 molecules, with 10~ps production runs, including some at low temperature (equilibrated at 260~K). Full details of all simulations can be found in Appendix~\ref{appendix-overview}, and are highlighted in the text where necessary.

\begin{figure}
\includegraphics{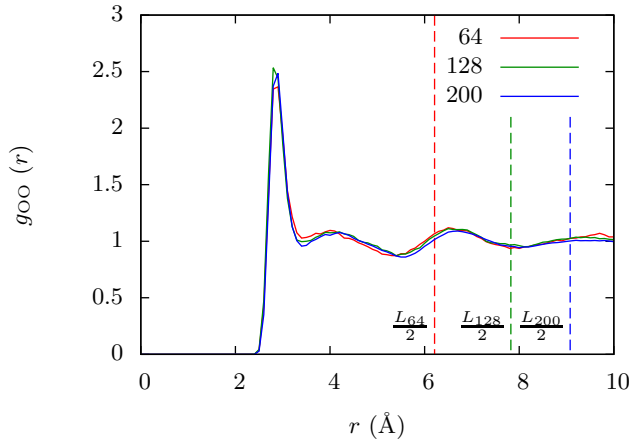}
\caption{Convergence of the O--O RDF with system size; the number of molecules in the simulation box $N_m$ is given in the key. The vertical dashed lines indicate half the box size $L_{N_m}$.}
\label{fig:rdf_finite}
\end{figure}

Fig.~\ref{fig:rdf_finite} shows the dependence on the size of the simulation box of the calculated O--O radial distribution function (RDF) at ambient conditions. Finite size errors are almost negligible, in agreement with previous studies~\cite{Fernandez-Serra2004,Kuhne2009,water_emiliomarivi}. Furthermore, the RDF for the 200-molecule box is found to be extremely stable for both xc functionals: when calculating it within a moving time window of 2.5~ps, we see no noticeable change throughout the entire 20~ps production run other than small fluctuations within the statistical error.

\subsection{Basis set tests}

We have extensively tested the accuracy of our \DZPmv\ basis, comparing it to a much larger quadruple-$\zeta$ doubly-polarized basis (\QZDP), as well as to fully converged plane-wave (PW) calculations. Our investigation of NAO basis sets for water systems are described in detail elsewhere~\cite{Corsetti2013}.

\begin{figure}
\includegraphics{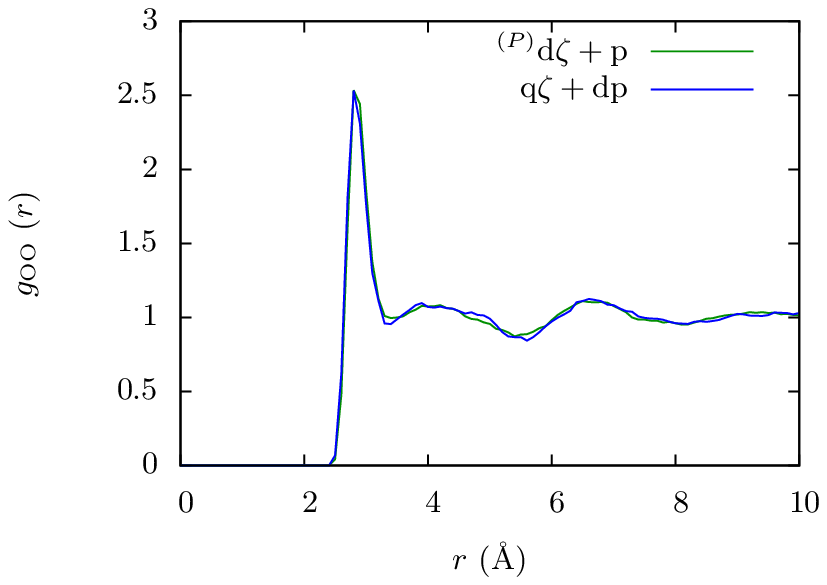}
\caption{Comparison of the O--O RDFs obtained using the standard \DZPmv\ and high-quality \QZDP\ bases.}
\label{fig:rdf_basis}
\end{figure}

Fig.~\ref{fig:rdf_basis} shows the comparison of the RDFs obtained using the two NAO bases. Due to the increased computational cost, the simulation using \QZDP\ was run for a shorter time (2.5~ps after 3~ps of equilibration), and, hence, the RDF is less smooth. Nevertheless, there is an excellent agreement between the two. Other quantities of interest require longer simulation times for an accurate measurement, and so the comparison should be taken with caution; the results obtained, however, suggest a good agreement for the self-diffusion coefficient (within the statistical error), and an underestimation of $\sim$3~kbar for the average pressure calculated with \DZPmv.

The comparison with PWs was carried out with the ABINIT~\cite{abinit-generic-et-al2} code. The same pseudopotentials are used in both codes, with identical Kleinman-Bylander factorizations~\cite{pseudo-KB}, and local and non-local components. For the tests, we make use of two sets of 100 uncorrelated snapshots of the liquid in a 32-molecule box (obtained by very long simulations with the TIP4P force field), one at 1.00~g/cm$^3$ and the other at 1.20~g/cm$^3$. PW calculations of these 200 snapshots are performed for a range of kinetic energy cutoffs, up to a very high cutoff (2700 eV) for which we can consider the results to be fully converged.

We have calculated the root mean square (RMS) error in the two test sets with respect to the converged PW results for several quantities that give a good indication of the level of accuracy of the NAO bases: total energy differences between snapshots, ionic forces, and absolute pressures. We find small RMS errors for \DZPmv\ of 1.7~meV/molecule in energy differences and 0.11~eV/\AA\ (O ions) and 0.07~eV/\AA\ (H ions) in the magnitude of the forces. The corresponding values for \QZDP\ are 0.5~meV/molecule in energy differences and 0.03~eV/\AA\ (O ions) and 0.02~eV/\AA\ (H ions) in the magnitude of the forces. In both cases the differences between the two test sets are negligible. When comparing with the RMS errors obtained for PW bases at different kinetic energy cutoffs, these results show \DZPmv\ to be comparable to the accuracy of a $\sim$850~eV cutoff, while \QZDP\ is comparable to a $\sim$1000~eV cutoff.

\begin{figure}
\includegraphics{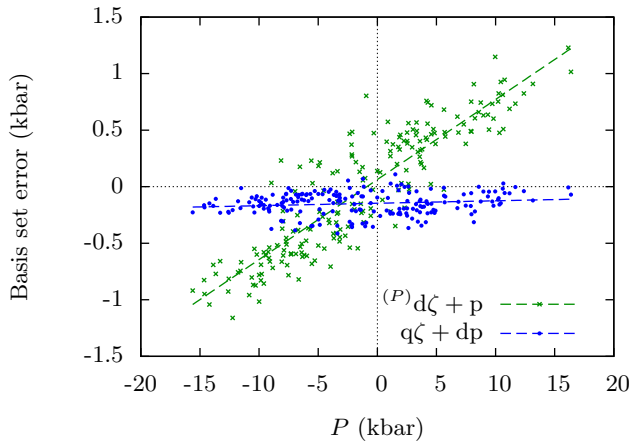}
\caption{Error in pressures calculated in SIESTA with respect to converged PW pressures for 200 snapshots of the liquid at two different densities. The dashed lines show the fits used for the basis set correction.}
\label{fig:P_conv}
\end{figure}

When calculating pressure values, the NAO bases show a noticeable advantage over PWs, as the latter suffer from a spurious tensile stress introduced by the effective change in kinetic energy cutoff associated with the infinitesimal change in volume. Even {\em after} correcting for this error in the PW values~\cite{Meade1989}, however, the two NAO bases give accuracies comparable to a $\sim$1500~eV cutoff when considering the RMS error in absolute pressures. Fig.~\ref{fig:P_conv} shows a scatter plot of the error in the pressure values calculated with the NAO bases against the fully-converged values, for all snapshots in both test sets. The error is fitted with a linear function in $P$; we use this fit to correct all pressures obtained from our AIMD simulations. The effect of the correction is small on the scale we are interested in (causing changes in average pressures 1--2 orders of magnitude smaller than the pressure range shown in Fig.~\ref{fig:EoS}). Furthermore, as our simulations are performed at fixed volume, errors in $P$ will not affect the AIMD trajectory.

\subsection{Non-local vdW density functionals}

For the vdW-DF functional, we substitute the revPBE~\cite{revPBE} exchange energy with PBE~\cite{pbe}, as previous studies have shown this to noticeably improve the calculated RDF~\cite{water_emiliomarivi,Zhang2011a} due to a better description of H bonds. We refer to the resulting functional as vdW-DF$^\mathrm{PBE}$. For VV10, we use PW86R~\cite{PW86RPBE} exchange, as suggested by its authors~\cite{VV10}.

The non-local correlation energy can be written as
\begin{equation}
E_c^{nl} = \frac{1}{2} \int\int \mathrm{d}^3\rr_1 ~\mathrm{d}^3\rr_2 ~n_1 ~n_2 ~\phi(n_1,|\nabla n_1|,n_2,|\nabla n_2|,r_{12}),
\label{eqn:Enl}
\end{equation}
where $n_i$ and $|\gn_i|$ are the electron density and its gradient at $\rr_i$, and $r_{12}=|\rr_1-\rr_2|$. In vdW-DF, the variables $n_i$ and $|\gn_i|$ are contracted in an auxiliary variable $q_i = q(n_i,|\gn_i|)$, and this was used by Rom\'{a}n-P\'{e}rez and Soler~\cite{RomanPerez2009} to approximate $\phi(q_1,q_2,r_{12})$ by an interpolation series in $q_1$ and $q_2$. This allows the use of the convolution theorem and fast Fourier transforms for each term of the series. In contrast, in VV10 the non-local kernel $\phi$ depends {\em independently} on the electron density and its gradient at each of $\rr_1$ and $\rr_2$. Consequently, it requires a four- rather than a two-dimensional interpolation and, in principle, many more interpolation points~\cite{Sabatini2013}. To address this problem, we use a proposal by Wu and Gygi~\cite{Wu-Gygi2012}: in order to handle the logarithmic singularity present in vdW-DF, they suggest not to interpolate $\phi$ itself, but the product $q_1 q_2 \phi$, which is smooth at $q_1 = q_2 = 0$. Similarly, we find that the whole integrand $n_1 n_2 \phi$ for VV10 is much smoother than $\phi$ alone. This means that as few as $(7 \times 5)$ points suffice for an accurate interpolation in $k_F = (3 \pi^2 n)^{1/3}$ and $k_G = |\gn|/n$. This is comparable to the $\sim 20$--$30$ points used originally to interpolate $\phi$ as a function of $q$. Thus, the cost of VV10 and vdW-DF becomes similar (both with a small overhead of $\sim$20\% relative to GGA functionals), thereby allowing AIMD simulations of large systems, as necessary for the present study.

\section{Results}

\subsection{Compressibility from the pressure--density curve}

\begin{figure}
\includegraphics{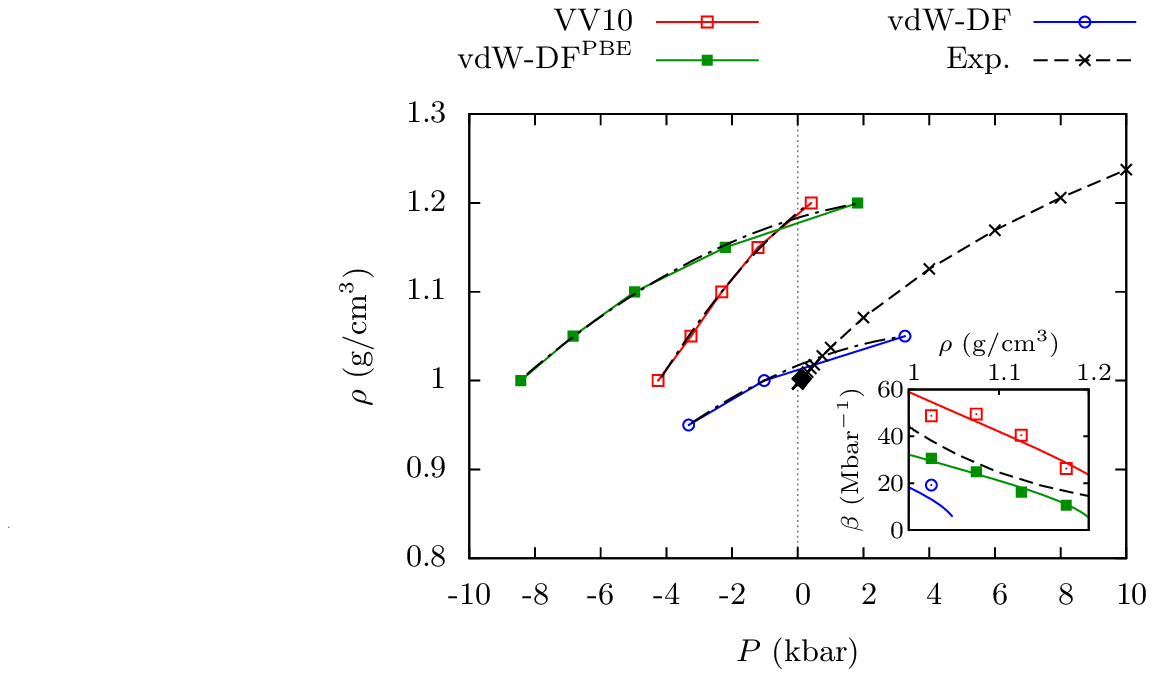}
\caption{Pressure--density curves from AIMD simulations at fixed density (200 molecules, $\sim$300~K). Experimental data at 300~K from Ref.~[\onlinecite{water_density}]; data for vdW-DF (64 molecules, $\sim$300~K) from Ref.~[\onlinecite{water_emiliomarivi}]. Black dashed--dotted lines show the fitted virial equation for each functional. The inset shows the corresponding compressibility--density curves (solid lines from the fit, and points from finite differences).}
\label{fig:EoS}
\end{figure}

\begin{figure}
\includegraphics{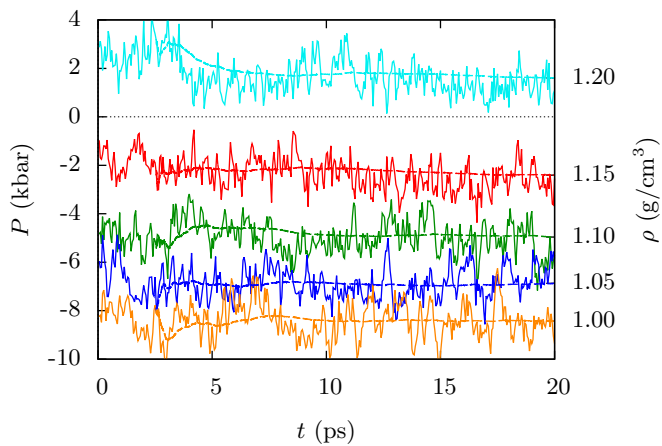}
\caption{Evolution of the pressure of the system during the production run for the vdW-DF$^\mathrm{PBE}$ AIMD simulations at different densities (200 molecules, $\sim$300~K). The solid lines show the instantaneous pressure averaged over intervals of 50~fs, and the dashed lines show the cumulative running average starting from $t=2.5$~ps.}
\label{fig:P_finite}
\end{figure}

The calculated $P$--$\rho$ equations of state for VV10 and vdW-DF$^\mathrm{PBE}$ are shown in Fig.~\ref{fig:EoS}, alongside previous results for vdW-DF (with revPBE exchange). We perform a series of AIMD simulations at fixed densities between 1.00 and 1.20~g/cm$^3$, in steps of 0.05~g/cm$^3$, and we fit $\rho(P)$ to a virial equation in powers of $P$, up to second order. The excellent fits indicate the small uncertainty of our average pressures. This is also shown in Fig.~\ref{fig:P_finite}: the cumulative running average for the instantaneous pressure stays almost constant after the first 10~ps of the simulation. We have carried out the same simulations for vdW-DF$^\mathrm{PBE}$ with a smaller cell of 128 molecules. The pressure difference between the two sizes is small (0.6~kbar RMS), confirming that we are well converged in system size, in agreement with previous tests performed with the TIP4P force field~\cite{water_emiliomarivi}. However, the error bars are noticeably larger for the smaller system, leading to a worse fit for $\rho(P)$.

As shown by Wang {\em et al.}~\cite{water_emiliomarivi}, the original vdW-DF functional gives the best density of liquid water at ambient pressure (1.00~g/cm$^3$ at $\sim$0.0~kbar), perfectly correcting for the low densities given by all GGA functionals. However, it severely underestimates the compressibility, suggesting that the agreement at ambient pressure is fortuitous. This is confirmed by its RDF at 1.00~g/cm$^3$, which is in much poorer agreement with experiment than even those obtained using GGA.

\begin{table*}
\caption{Compressibility of liquid water calculated for different vdW xc functionals, both at the experimental ambient density and ambient pressure. Errors with respect to experiment are given in brackets.}
\label{table:comp}
\begin{ruledtabular}
{\footnotesize
\begin{tabular*}{\textwidth}{lcccc}
& Exp.~\cite{water_density} & VV10 (error) & vdW-DF$^\mathrm{PBE}$ (error) & vdW-DF~\cite{water_emiliomarivi} (error) \\
\hline
$\beta$ ($\rho=1.00$~g/cm$^3$) (Mbar$^{-1}$) & \multirow{2}{*}{45.0} & 59.0 (+31\%)   & 32.2 ($-$28\%)                & 18.2 ($-$60\%) \\
$\beta$ ($P=0.0$~kbar) (Mbar$^{-1}$)         &                       & 25.8 ($-$43\%) & {\color{white}0}9.4 ($-$79\%) & 14.8 ($-$67\%)
\end{tabular*}
}
\end{ruledtabular}
\end{table*}

In contrast, our results show that VV10 and vdW-DF$^\mathrm{PBE}$ reproduce the overall shape of the equation of state much better, despite a shift towards negative pressures that results in an equilibrium ambient density overestimated by almost 20\% in both cases. Our results for the compressibility $\beta=(1/\rho)(\partial \rho / \partial P)$ are given in Table~\ref{table:comp} and in the inset of Fig.~\ref{fig:EoS}. The experimental value is between those of VV10 and vdW-DF$^\mathrm{PBE}$ over the entire density range, with discrepancies of $\sim$30\% at 1.00~g/cm$^3$, while vdW-DF greatly underestimates it.

It is interesting to compare these results with those reported by Pi {\em et al.}~\cite{Vega2009} for four popular force-field models, all of which overestimate the compressibility at ambient density, with errors ranging from 26.7\% (for TIP5P~\cite{TIP5P}) down to only 2.4\% for SPC/E~\cite{SPCE} and 2.9\% for TIP4P/2005~\cite{Abascal2005}. However, when examining the change in compressibility with temperature, it is clear that only one model, TIP4P/2005, correctly reproduces the experimental data for a wide range of temperatures (including the existence of a minimum at $\sim$320~K), while the small error at ambient temperature for SPC/E can be seen to be fortuitous. Furthermore, TIP5P does not show any sign of the anomalous behaviour in the range considered. It is encouraging to note that, when analysing our results for vdW-DF$^\mathrm{PBE}$ with 128 molecules at two different temperatures ($\sim$260~K and $\sim$300~K, as listed in Appendix~\ref{appendix-overview}) with the best fit through the data points, we find an increase in the compressibility at 1.00~g/cm$^3$ of $\sim$20\% for the low temperature simulations respect to the ambient temperature ones, in reasonable agreement with experiment~\cite{Vega2009} (27.6\%), and better than TIP4P/2005 (11.0\%). It seems likely, therefore, that vdW-DF$^\mathrm{PBE}$ will also exhibit the compressibility anomaly.

\subsection{Structural variations with density}

\begin{figure}
\includegraphics{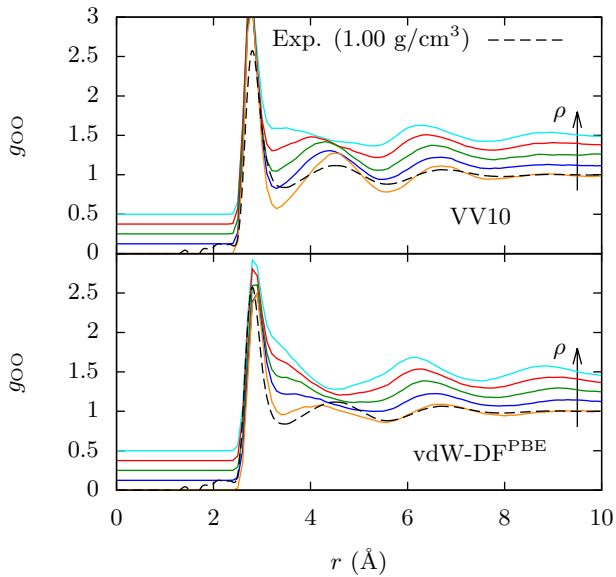}
\caption{Structural variations with density from AIMD simulations (1.00~g/cm$^3$ to 1.20~g/cm$^3$, 200 molecules, $\sim$300 K). The dashed black line shows the experimental data at ambient conditions from Ref.~[\onlinecite{waterRDF2013}], corresponding to the lowest AIMD density (the solid orange line).}
\label{fig:rdf_trends}
\end{figure}

\begin{figure}
\includegraphics{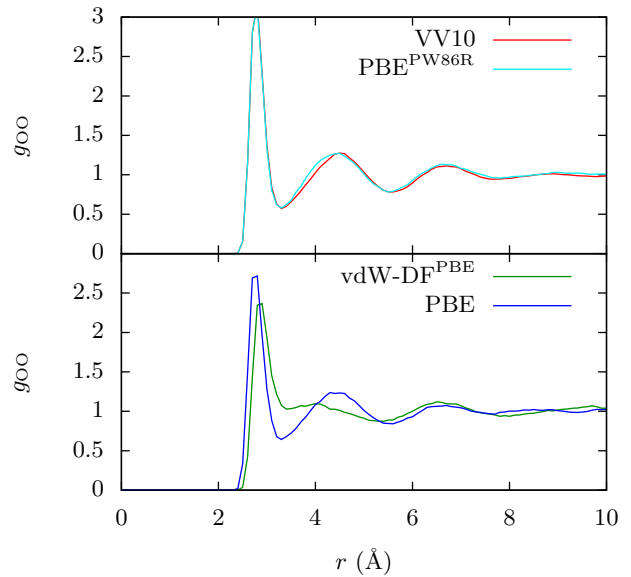}
\caption{Comparison of the O--O RDFs (1.00~g/cm$^3$, $\sim$300~K) obtained with a vdW xc functional and with the underlying semi-local functional it uses. In the upper panel, the VV10 and PBE$^\mathrm{PW86R}$ simulations are of 200 and 128 molecules, respectively; in the lower panel, both the vdW-DF$^\mathrm{PBE}$ and PBE simulations are of 64 molecules. PBE data from Ref.~[\onlinecite{water_emiliomarivi}].}
\label{fig:rdf_PW86RPBEPBE}
\end{figure}

The RDFs of VV10 and vdW-DF$^\mathrm{PBE}$ are compared with a recent determination from x-ray diffraction~\cite{waterRDF2013} at ambient conditions (the lowest density line in Fig.~\ref{fig:rdf_trends}). VV10 is overstructured, but with excellent agreement in the position of the extrema, while vdW-DF$^\mathrm{PBE}$ is only slightly understructured, with a small outwards shift of $\sim$0.1~\AA\ for the first maximum and a larger inwards shift of $\sim$0.3~\AA\ for the second one. As explained by Wang {\em et al.}, vdW-DF$^\mathrm{PBE}$ corrects the low density of the GGA liquid by favoring the population of the anti-tetrahedral interstitial sites of the H-bond network, at the cost of breaking some H bonds. This results in a significantly different RDF for vdW-DF$^\mathrm{PBE}$ respect to its underlying PBE functional (lower panel of Fig.~\ref{fig:rdf_PW86RPBEPBE}). These anti-tetrahedral interstitial sites correspond to vdW-induced local minima in non-H-bonded dimer configurations.

\begin{figure}
\includegraphics{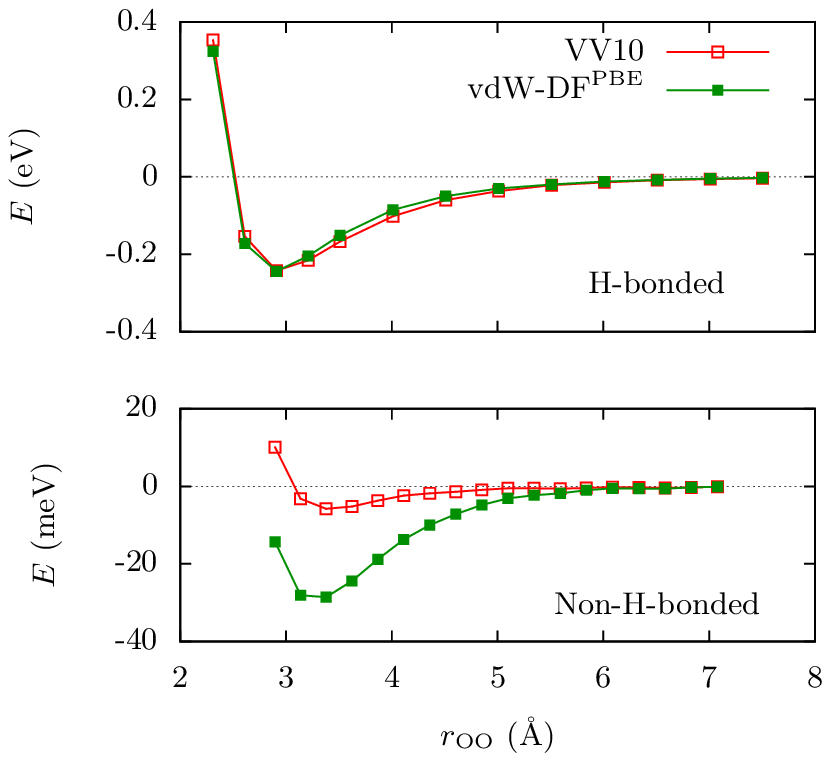}
\caption{Total energy of the water dimer as a function of the intermolecular separation $r_\mathrm{OO}$ in the H-bonded and non-H-bonded configurations. The molecular orientations are the same as those pictured in Fig.~7 of Ref.~[\onlinecite{water_emiliomarivi}]. We note the different energy scales for the two configurations.}
\label{fig:dimer}
\end{figure}

We have compared the energetics of H-bonded and non-H-bonded configurations for VV10 and vdW-DF$^\mathrm{PBE}$ (Fig.~\ref{fig:dimer}). While the H-bonded binding energy is almost identical for both functionals (247~meV for VV10 and 245~meV for vdW-DF$^\mathrm{PBE}$), the non-H-bonded one is almost five times stronger in vdW-DF$^\mathrm{PBE}$ (29~meV) than in VV10 (6~meV). 

We have seen that the density of the VV10 liquid increases with respect to that of GGA functionals. In the upper panel of Fig.~\ref{fig:rdf_PW86RPBEPBE}, we show the O--O RDFs for VV10 and PBE$^\mathrm{PW86R}$, the underlying semi-local functional used in VV10 (see Table~\ref{apptable:xc_defs}), both at the same density of 1.00~g/cm$^3$. Despite the two RDFs being very similar in this case, the corresponding pressures are very different ($-$4.2~kbar and 3.5~kbar, respectively). Therefore, the reason why the density of VV10 water increases with respect to its underlying semi-local functional is due to a different mechanism to the one described above for vdW-DF$^\mathrm{PBE}$. While the topology of the H-bond network is not modified, there is an overall attraction between non-H-bonded molecules that reduces the pressure of the simulation. In vdW-DF$^\mathrm{PBE}$ the non-H-bonded binding energy is so large that it favors the breaking of weak H bonds, modifying the structure of the H-bond network and favoring the positioning of molecules at anti-tetrahedral interstitial sites.

\begin{figure}
\includegraphics{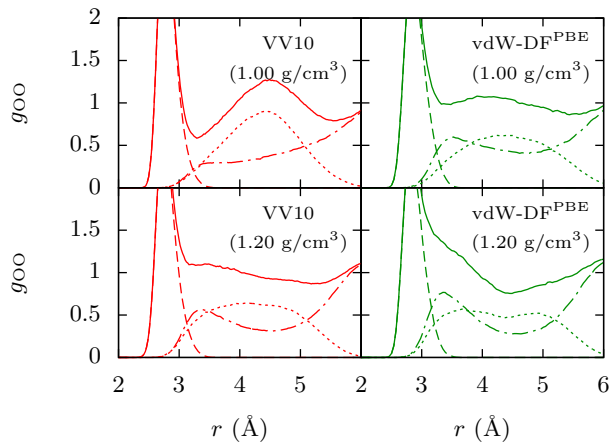}
\caption{Decomposition of the O--O RDFs at two different densities (200 molecules, $\sim$300~K). Solid lines show the total RDF, long-dashed lines the first H-bonded shell, short-dashed lines the second H-bonded shell, and dashed--dotted lines the remaining molecules (including the non-H-bonded interstitial shell).}
\label{fig:rdf_decomp}
\end{figure}

Fig.~\ref{fig:rdf_trends} also shows the change of the RDFs with increasing density, from 1.00~g/cm$^3$ to 1.20~g/cm$^3$ (see Appendix~\ref{appendix-overview} for more details on the behavior of the extrema). Of these, Fig.~\ref{fig:rdf_decomp} shows the lowest and highest densities only, decomposed in terms of the H-bond network (see Appendix~\ref{appendix-HB} for a description of our H-bond definition). We see a different behavior for the two functionals. Although in both cases the liquid becomes less structured at higher density, for VV10 the second H-bonded shell moves inwards, closing the $\widehat{\mathrm{OOO}}$ angle. For vdW-DF$^\mathrm{PBE}$ the increase in pressure induces a larger population of the interstitial anti-tetrahedral sites, and the second H-bonded shell presents a bimodal behavior, some molecules moving inwards and some outwards (opening the $\widehat{\mathrm{OOO}}$ angle).

\begin{figure}
\includegraphics{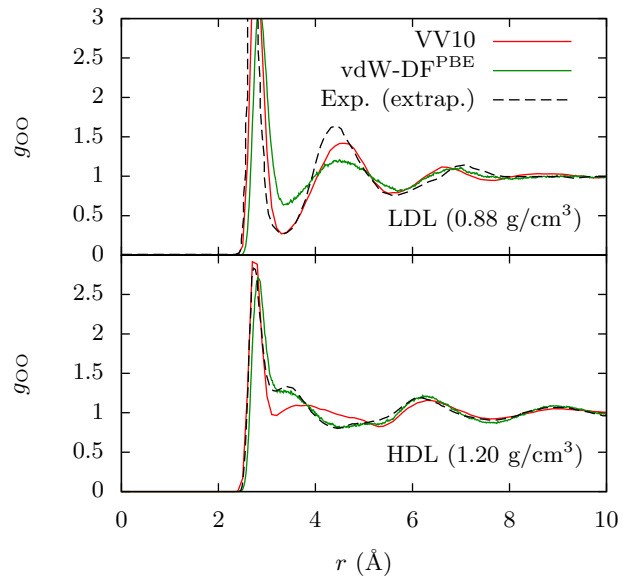}
\caption{Comparison of the low-temperature O--O RDFs from AIMD simulations (128 molecules, $\sim$260~K) with experiment at 268~K in the low- and high-density limits. Experimental RDFs are obtained from a linear extrapolation of several intermediate densities~\cite{soper}.}
\label{fig:rdf_LHDL}
\end{figure}

These results can be compared with those reported by Soper and Ricci~\cite{soper} for neutron diffraction of cold water under pressure. They assume that the measured structure factors can be described as a linear combination of two components, corresponding to a low- and a high-density liquid (LDL and HDL, respectively). Fitting their data to this model for a range of pressures, they extrapolate to obtain the pure LDL and HDL structures and densities at 268~K. We have simulated 128 molecules at the predicted densities for LDL and HDL, at $\sim$260~K. Fig~\ref{fig:rdf_LHDL} compares the resulting {\em ab initio} RDFs with the experimentally extrapolated ones. There is a very good agreement between low-density VV10 and the predicted LDL, and a truly remarkable agreement between high-density vdW-DF$^\mathrm{PBE}$ and the predicted HDL. In particular, vdW-DF$^\mathrm{PBE}$ reproduces an extended minimum at $r \simeq 4.5$~\AA\ due to the bimodal structure of the second H-bonded shell~\footnote{Interestingly, a recent classical force-field simulation~\cite{Kaya2013} of thin-film water on a BaF$_2$ surface has reported a very similar extended minimum in the RDF for the first 1~\AA-thick water layer above the surface, which is therefore suggested to be in a highly compressed state similar to HDL.}. On the other hand, the two functionals show quite different features for the opposite limits: vdW-DF$^\mathrm{PBE}$ is understructured at low density, while VV10 does not exhibit the correct structure at high density. Our findings, therefore, rationalize the discrepancies observed at intermediate densities, for which both functionals give too large a weight to one of the two components~\footnote{The $P$--$\rho$ curves in Fig.~\ref{fig:EoS} arguably reveal the same behavior: by extrapolating the vdW-DF$^\mathrm{PBE}$ curve, we can see a tentative crossing with experiment around 1.25~g/cm$^3$ and, analogously, a crossing of VV10 with experiment at low density (the precise value being harder to estimate in this case).}.

\begin{figure*}
\includegraphics{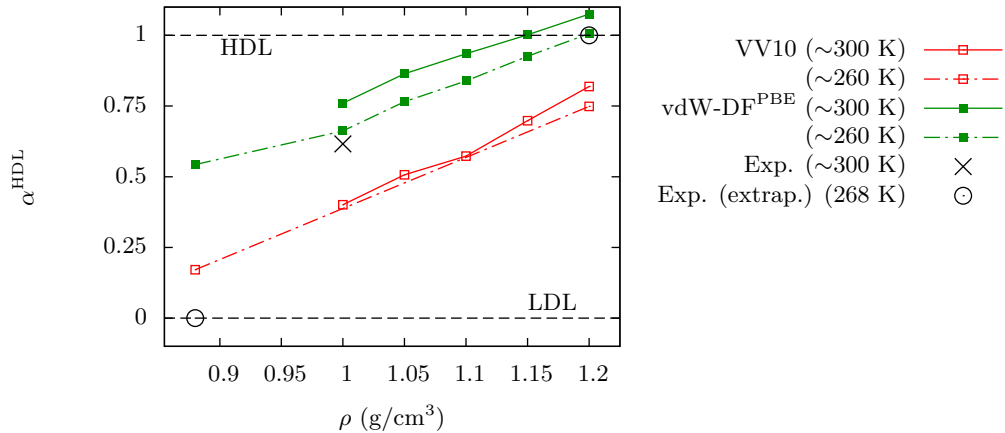}
\caption{Best fit value for the mixing parameter $\alpha^\mathrm{HDL}$ in the LDL/HDL linear mixing model, for the O--O RDFs from AIMD simulations. The values at $\sim$300~K are for simulations of 200 molecules, and those at $\sim$260~K of 128 molecules. The two open circles give the end points of the mixing~\cite{soper}, while the cross gives the best fit value for the experimental RDF at ambient conditions from Ref.~[\onlinecite{waterRDF2013}].}
\label{fig:mix}
\end{figure*}

We now investigate this conclusion in more detail, by attempting a quantitative analysis based on the same linear mixing model employed by Soper and Ricci. The RDF at a given temperature and density (and for a given xc functional) is assumed to be a simple linear combination of the HDL and LDL RDFs extrapolated from experimental data, with weights of $\alpha^\mathrm{HDL}$ and $\alpha^\mathrm{LDL}=1-\alpha^\mathrm{HDL}$, respectively. We neglect cross terms between the two RDFs~\cite{Mogelhoj2011}, as well as any additional dependence of each individual RDF on temperature or pressure. For any particular RDF obtained by AIMD simulation, we can then find the `best fit' value of $\alpha^\mathrm{HDL}$ by minimizing the quantity
\begin{equation}
Q \left ( \alpha^\mathrm{HDL} \right ) =\int_{r_1}^{r_2} \left | g_\mathrm{OO}^\mathrm{AIMD} \left ( r \right ) - g_\mathrm{OO}^\mathrm{mix} \left ( r; \alpha^\mathrm{HDL} \right ) \right |^2 \mathrm{d}r,
\end{equation}
where we take $r_1=3$~\AA\ and $r_2=10$~\AA\ (the first peak is excluded, since small variations in its width can result in large changes in height, thereby swamping the more relevant and reliable variations around the second peak). The results of this analysis are shown in Fig~\ref{fig:mix}. The previously postulated behavior of the two functionals can now be seen very clearly: vdW-DF$^\mathrm{PBE}$ accurately recovers the HDL structure at high density, while VV10 fairly accurately recovers the LDL one at low density. However, in both cases the rate of change of $\alpha^\mathrm{HDL}$ with density is too small, resulting in each functional giving too much weight to its preferred structure at the opposite limit, as well as in between (e.g., at ambient density). We note that we find $\alpha^\mathrm{HDL}>1$ for vdW-DF$^\mathrm{PBE}$ at ambient temperature and high density; this is a possible indication that the extrapolated end-point structures are themselves not pure HDL or LDL, but still contain a small amount of mixing.

It is interesting to note that our results confirm the suggestion of M\o{}gelh\o{}j {\em et al.}~\cite{Mogelhoj2011} of the similarity of the RDF obtained using a semi-local functional (PBE) with the LDL structure, and of those obtained using two non-local functionals (optPBE-vdW~\cite{Klimes2010} and vdW-DF2~\cite{Lee2010}) with the HDL structure. Their simulations are performed at ambient temperature and density, and, indeed, we find a good agreement between the RDF for PBE and that for VV10 (also at ambient temperature and density), and a fairly good agreement between the RDFs for the two non-local functionals and that for vdW-DF$^\mathrm{PBE}$. As we have discussed, VV10 barely modifies the RDF of its underlying semi-local functional, hence the similarity with PBE as well. The similarity between vdW-DF$^\mathrm{PBE}$, optPBE-vdW, and vdW-DF2 is also not surprising, as they are all related to the vdW-DF functional of Dion {\em et al.}~\cite{vdW-DF}, with some modifications in each case that improve the H-bond description. However, the latter two are even more severely understructured than vdW-DF$^\mathrm{PBE}$ at ambient conditions (this discrepancy is reduced but not eliminated by accounting for finite size effects, shown in Fig.~\ref{fig:rdf_finite} for vdW-DF$^\mathrm{PBE}$).

\subsection{Diffusivity}

Finally, we show that our simulations allow us to evaluate one of the anomalies of the liquid. The experimental self-diffusion coefficient $D$ increases with density (or pressure) up to a maximum at 1.03~g/cm$^3$ (at $\sim$300~K), above which the expected decrease is observed~\cite{H20_diff_max,H2O_D2O_diff}. This anomaly has been thoroughly studied and is well reproduced by empirical force fields~\cite{Scala2000,Errington2001,Netz2002}. However, first principles calculations of $D$ are especially challenging due to the relatively small system sizes and simulation times accessible with AIMD~\cite{Fernandez-Serra2004,Kuo2004}; in particular, system size effects are much larger for dynamical properties than for structural ones~\cite{Dunweg1993,Kuhne2009}.

\begin{figure}
\includegraphics{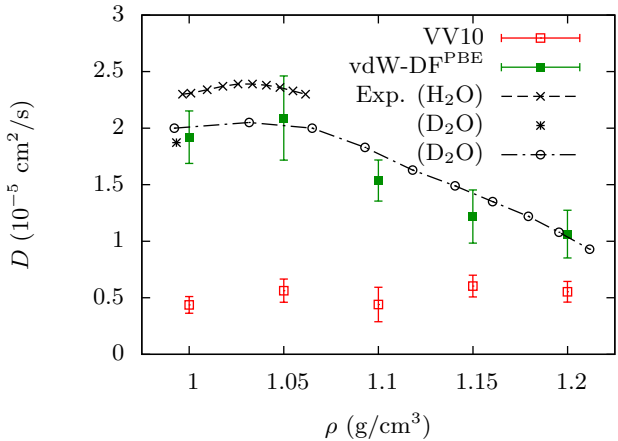}
\caption{Variation of the self-diffusion coefficient with density from AIMD simulations (200 D$_2$O molecules, $\sim$300~K). Experimental data for H$_2$O from Ref.~[\onlinecite{H20_diff_max}] (298~K), and for D$_2$O from Ref.~[\onlinecite{H2O_D2O_diff}] (single point, 298~K) and Ref.~[\onlinecite{D20_diff_max}] (trend, 303~K).}
\label{fig:diff}
\end{figure}

Our large 200-molecule simulation box and reasonably long AIMD runs have reduced the uncertainty enough to observe a clear trend in $D (\rho)$, as shown in Fig.~\ref{fig:diff}. The finite size correction proposed by D\"{u}nweg and Kremer~\cite{Dunweg1993} (not included) is fairly small for our system size: an increase of $\sim$$0.16 \times 10^{-5}$~cm$^2$/s using the parametrization calculated by K\"{u}hne {\em et al.} for PBE~\cite{Kuhne2009}.

We find vdW-DF$^\mathrm{PBE}$ to be extremely successful, perfectly reproducing NMR spin-echo measurements~\cite{D20_diff_max} within the statistical errors. Furthermore, within our density resolution, $D (\rho)$ shows a maximum at 1.05~g/cm$^3$, in agreement with experiment. However, this result must be carefully assessed, as the density point giving the diffusivity maximum equilibrated to a temperature $\sim$10~K higher than the two points around it. This suggests that it might only be an apparent maximum, caused by the increased temperature at that point (in fact, this is also reflected in the large error bars). Indeed, experimental results show that an increase of 10~K in temperature can cause an increase of 0.5--0.6~g/cm$^3$ in the self-diffusion coefficient both for light and heavy water~\cite{H20_diff_max,H2O_D2O_diff,D20_diff_max}. On the other hand, it also possible that the diffusivity of the liquid is mostly determined by the original temperature (and the corresponding structure) at which the system was equilibrated (300~K for all points), rather than the average temperature reached during the NVE simulation. In favor of this latter hypothesis, we note that the simulations at 1.15~g/cm$^3$ and 1.20~g/cm$^3$ also featured a similar increase in temperature; however, applying an approximate correction only to these three points results in a less smooth trend overall, with a minimum at 1.05~g/cm$^3$ and a maximum at 1.10~g/cm$^3$. Furthermore, using this correction, the only points to maintain a good agreement with the experimental curve would be those at 1.00~g/cm$^3$ and 1.10~g/cm$^3$ (i.e., the uncorrected ones). It seems quite unlikely that, while these two points do indeed agree with experiment, the equally good agreement of the other three points shown in Fig.~\ref{fig:diff} is merely fortuitous.

In contrast to these promising results, VV10 shows a nearly density-independent and significantly underestimated diffusivity (by 78\% at 1.00~g/cm$^3$, similarly to previous GGA results~\cite{Fernandez-Serra2004}). This is directly related to an overstructured liquid~\cite{Fernandez-Serra2004,Kuo2004}: for all the densities considered, $D (\rho)$ indicates that the system is effectively supercooled, with too strong a H-bond network.

\section{Conclusions}

We have performed a detailed study of first principles models of liquid water with DFT, using two non-local density functionals that describe vdW interactions without empirical parameters. We have calculated the $P$--$\rho$ equation of state at room temperature, from which we extract the equilibrium ambient density and the compressibility, structural properties given by the O--O RDF, and the diffusivity as a function of density.

For these properties, we find that the vdW-DF functional, with PBE exchange, arguably gives the better description of the liquid. In particular, the self-diffusion coefficient is in excellent agreement with experiments and it appears to correctly reproduce the isothermal anomaly, within the error margins of the AIMD simulations. The differences between xc functionals also provide valuable insights. Interestingly, vdW-DF$^\mathrm{PBE}$ and VV10 seem to complement each other, by describing respectively the high- and low-density structures of water with remarkable precision. Thus, to reach a better description of the liquid, new functionals should improve the energy landscape between these two structures.

\begin{acknowledgments}
This work was partly funded by grants FIS2009-12721 and FIS2012-37549 from the Spanish Ministry of Science. MVFS acknowledges a DOE Early Career Award No. DE-SC0003871. The calculations were performed on the following HPC clusters: kroketa (CIC nanoGUNE, Spain), arina (Universidad del Pa\'{i}s Vasco/Euskal Herriko Unibertsitatea, Spain), magerit (CeSViMa, Universidad Polit\'{e}cnica de Madrid, Spain). We thank the RES--Red Espa\~nola de Supercomputaci\'{o}n for access to magerit. SGIker (UPV/EHU, MICINN, GV/EJ, ERDF and ESF) support is gratefully acknowledged.
\end{acknowledgments}

\appendix

\section{Overview of simulations}
\label{appendix-overview}

\begin{table*}
\caption{Specifications for the various density functionals considered: for each we list the exchange, (semi-)local correlation, and non-local correlation components. Both for VV10 and VV10$^\mathrm{revPBE}$, we use the parameters $b=5.9$ and $C=0.0093$ given in Ref.~[\onlinecite{VV10}].}
\label{apptable:xc_defs}
\begin{ruledtabular}
{\footnotesize \begin{tabular*}{\textwidth}{l c c c c c}
$E_\mathrm{xc}$            & vdW-DF               & vdW-DF$^\mathrm{PBE}$ & PBE$^\mathrm{PW86R}$  & VV10             & VV10$^\mathrm{revPBE}$ \\
\hline                                                                    
$E_\mathrm{x}$             & revPBE~\cite{revPBE} & PBE~\cite{pbe}        & PW86R~\cite{PW86RPBE} & PW86R            & revPBE                 \\
$E_\mathrm{c}^0$           & LDA~\cite{qmc2}      & LDA                   & PBE                   & PBE              & PBE                    \\
$E_\mathrm{c}^\mathrm{nl}$ & vdW-DF~\cite{vdW-DF} & vdW-DF                & -                     & VV10~\cite{VV10} & VV10
\end{tabular*}}
\end{ruledtabular}
\end{table*}

\begin{table*}
\caption{Details of all AIMD simulations. Listed are the number of molecules in the simulation box ($N_m$), the xc functional ($E_\mathrm{xc}$), the duration of the production run ($\tau_\mathrm{run}$), the density ($\rho$), the average and target temperatures ($T_\mathrm{av}$ and $T_\mathrm{equil}$, respectively), the average pressure ($P_\mathrm{av}$) and the pressure estimate at the target temperature ($P_\mathrm{corr}$), the average self-diffusion coefficient ($D_\mathrm{av}$), and the coordinates of the first two maxima and minima in the RDF ($r^{(i)}$, $g_\mathrm{OO}^{(i)}$).}
\label{apptable:all_runs}
\begin{ruledtabular}
{\scriptsize \begin{tabular*}{\textwidth}{l c c c c c c c c c c}
\multirow{2}{*}{$N_\mathrm{m}$} & \multirow{2}{*}{$E_\mathrm{xc}$} & $\tau_\mathrm{run}$ & $\rho$                       & $T_\mathrm{av}$ ($T_\mathrm{equil}$) & $P_\mathrm{av}$ ($P_\mathrm{corr}$) & $D_\mathrm{av}$               & \multicolumn{4}{c}{RDF extrema ($r^{(i)}$ (\AA), $g_\mathrm{OO}^{(i)}$)} \\ \cline{8-11}
                                &                                  & (ps)                & ($\mathrm{g}/\mathrm{cm}^3$) & (K)                                  & (kbar)                              & ($10^{-5}$ $\mathrm{cm}^2$/s) & 1$^\mathrm{st}$ max. & 1$^\mathrm{st}$ min. & 2$^\mathrm{nd}$ max. & 2$^\mathrm{nd}$ min. \\
\hline
\multirow{2}{*}{64}   & vdW-DF$^\mathrm{PBE}$                   & 10.0 & 1.00 & 308 (300) & $-$8.5 ($-$8.6) & 2.1 & $(2.9, 2.37)$ & $(3.5, 1.03)$ & $(4.0, 1.10)$ & $(5.4, 0.87)$ \\ \cline{2-11}
                      & vdW-DF                                  & 10.0 & 1.00 & 308 (300) &    0.7    (0.9) & 3.3 & $(2.9, 2.09)$ & -             & -             & $(4.9, 0.84)$ \\
\hline
\multirow{17}{*}{128} & \multirow{12}{*}{vdW-DF$^\mathrm{PBE}$} & 10.0 & 1.20 & 301 (300) &    1.1    (1.1) & 0.9 & $(2.8, 2.46)$ & -             & -             & $(4.4, 0.76)$ \\
                      &                                         & 10.0 & 1.15 & 316 (300) & $-$1.3 ($-$1.1) & 1.3 & $(2.8, 2.43)$ & -             & -             & $(4.6, 0.84)$ \\
                      &                                         & 10.0 & 1.10 & 309 (300) & $-$4.8 ($-$4.8) & 1.5 & $(2.9, 2.34)$ & -             & -             & $(5.1, 0.86)$ \\
                      &                                         & 10.0 & 1.05 & 304 (300) & $-$6.9 ($-$6.9) & 1.2 & $(2.8, 2.42)$ & $(3.4, 1.06)$ & $(3.8, 1.10)$ & $(5.5, 0.87)$ \\
                      &                                         & 10.0 & 1.00 & 301 (300) & $-$8.3 ($-$8.3) & 1.7 & $(2.8, 2.51)$ & $(3.4, 1.00)$ & $(3.9, 1.08)$ & $(5.4, 0.87)$ \\
                      &                                         & 10.0 & 0.88 & 312 (300) & $-$7.0 ($-$7.1) & 2.6 & $(2.9, 2.62)$ & $(3.5, 1.07)$ & $(3.8, 1.09)$ & $(5.5, 0.92)$ \\
                      &                                         & 10.0 & 1.20 & 265 (260) &    0.4    (0.3) & 0.5 & $(2.8, 2.66)$ & -             & -             & $(4.5, 0.82)$ \\
                      &                                         & 10.0 & 1.15 & 263 (260) & $-$3.4 ($-$3.4) & 0.4 & $(2.8, 2.69)$ & $(3.3, 1.17)$ & $(3.5, 1.22)$ & $(5.4, 0.85)$ \\
                      &                                         & 10.0 & 1.10 & 264 (260) & $-$5.6 ($-$5.6) & 0.6 & $(2.8, 2.55)$ & $(3.3, 1.06)$ & $(3.7, 1.13)$ & $(5.4, 0.82)$ \\
                      &                                         & 10.0 & 1.05 & 264 (260) & $-$8.1 ($-$8.1) & 0.5 & $(2.8, 2.63)$ & $(3.4, 0.94)$ & $(4.1, 1.11)$ & $(5.5, 0.82)$ \\
                      &                                         & 10.0 & 1.00 & 262 (260) & $-$8.9 ($-$9.0) & 0.6 & $(2.8, 2.75)$ & $(3.3, 0.83)$ & $(4.1, 1.15)$ & $(5.6, 0.82)$ \\
                      &                                         & 10.0 & 0.88 & 265 (260) & $-$9.1 ($-$9.3) & 0.8 & $(2.8, 2.99)$ & $(3.4, 0.65)$ & $(4.4, 1.20)$ & $(5.7, 0.82)$ \\ \cline{2-11}
& vdW-DF$^\mathrm{PBE}$           & \multirow{2}{*}{{\color{white}0}2.5} & \multirow{2}{*}{1.00} & \multirow{2}{*}{289 (300)} & \multirow{2}{*}{$-$5.2 ($-$5.0)} & \multirow{2}{*}{1.9} & \multirow{2}{*}{$(2.8, 2.53)$} & \multirow{2}{*}{$(3.4, 0.95)$} & \multirow{2}{*}{$(3.9, 1.10)$} & \multirow{2}{*}{$(5.6, 0.84)$} \\
& (\QZDP)                         &                                      &                       &                            &                                  &                      &                                &                                &                                &                                \\ \cline{2-11}
                      & PBE$^\mathrm{PW86R}$                    & 10.0 & 1.00 & 307 (300) &    3.8    (3.5) & 0.8 & $(2.8, 3.12)$ & $(3.3, 0.59)$ & $(4.5, 1.26)$ & $(5.5, 0.79)$ \\ \cline{2-11}
                      & \multirow{2}{*}{VV10}                   & 10.0 & 1.20 & 262 (260) &    0.6    (0.5) & 0.3 & $(2.7, 2.92)$ & $(3.2, 0.96)$ & $(3.6, 1.09)$ & $(5.3, 0.82)$ \\
                      &                                         & 10.0 & 0.88 & 267 (260) & $-$5.8 ($-$6.1) & 0.2 & $(2.8, 3.96)$ & $(3.3, 0.27)$ & $(4.6, 1.42)$ & $(5.5, 0.78)$ \\ \cline{2-11}
                      & VV10$^\mathrm{revPBE}$                  & 10.0 & 1.00 & 300 (300) &    1.5    (1.5) & 0.8 & $(2.8, 2.74)$ & $(3.3, 0.76)$ & $(4.3, 1.22)$ & $(5.6, 0.81)$ \\                
\hline                                                                    
\multirow{10}{*}{200} & \multirow{5}{*}{vdW-DF$^\mathrm{PBE}$}  & 20.0 & 1.20 & 308 (300) &    1.7    (1.8) & 1.1 & $(2.8, 2.42)$ & -             & -             & $(4.5, 0.78)$ \\
                      &                                         & 20.0 & 1.15 & 307 (300) & $-$2.3 ($-$2.2) & 1.2 & $(2.8, 2.42)$ & -             & -             & $(4.6, 0.83)$ \\
                      &                                         & 20.0 & 1.10 & 301 (300) & $-$5.0 ($-$5.0) & 1.5 & $(2.9, 2.35)$ & -             & -             & $(4.9, 0.86)$ \\
                      &                                         & 20.0 & 1.05 & 312 (300) & $-$6.8 ($-$6.8) & 2.1 & $(2.9, 2.36)$ & $(3.5, 1.10)$ & $(3.7, 1.10)$ & $(5.3, 0.87)$ \\
                      &                                         & 20.0 & 1.00 & 301 (300) & $-$8.4 ($-$8.4) & 1.9 & $(2.9, 2.47)$ & $(3.4, 0.96)$ & $(4.2, 1.08)$ & $(5.6, 0.86)$ \\ \cline{2-11}
                      & \multirow{5}{*}{VV10}                   & 20.0 & 1.20 & 304 (300) &    0.5    (0.4) & 0.6 & $(2.8, 2.74)$ & $(3.3, 1.09)$ & $(3.5, 1.10)$ & $(5.3, 0.87)$ \\
                      &                                         & 20.0 & 1.15 & 308 (300) & $-$0.8 ($-$1.2) & 0.6 & $(2.8, 2.81)$ & $(3.2, 0.94)$ & $(4.0, 1.10)$ & $(5.4, 0.85)$ \\
                      &                                         & 20.0 & 1.10 & 311 (300) & $-$2.2 ($-$2.3) & 0.4 & $(2.8, 2.93)$ & $(3.2, 0.78)$ & $(4.2, 1.14)$ & $(5.5, 0.81)$ \\
                      &                                         & 20.0 & 1.05 & 311 (300) & $-$3.1 ($-$3.2) & 0.6 & $(2.8, 3.04)$ & $(3.3, 0.71)$ & $(4.4, 1.18)$ & $(5.5, 0.82)$ \\
                      &                                         & 20.0 & 1.00 & 314 (300) & $-$3.4 ($-$4.2) & 0.4 & $(2.8, 3.23)$ & $(3.3, 0.57)$ & $(4.5, 1.25)$ & $(5.6, 0.79)$ \\
\hline \hline
\multicolumn{3}{c}{\multirow{2}{*}{Exp.~\cite{water_density,H2O_D2O_diff,H20_diff_max,waterRDF2013}}} & \multirow{2}{*}{1.00} & \multirow{2}{*}{$\sim$300} & \multirow{2}{*}{0.0} & 2.3 (H$_2$O) & \multirow{2}{*}{$(2.8, 2.58)$} & \multirow{2}{*}{$(3.5, 0.84)$} & \multirow{2}{*}{$(4.5, 1.12)$} & \multirow{2}{*}{$(5.6, 0.88)$} \\
                                                                                                    &&&                       &                            &                      & 1.9 (D$_2$O) &                                &                                &                                &
\end{tabular*}}
\end{ruledtabular}
\end{table*}

We have performed AIMD simulations at three different system sizes (64, 128 and 200 molecules), and using five different xc functionals (listed in Table~\ref{apptable:xc_defs}). Table~\ref{apptable:all_runs} details all these simulations, including both the input parameters and the post-processing results. Unless otherwise stated, we use the \DZPmv\ basis.

\begin{figure*}
\subfigure[\ Functionals based on VV10]{\includegraphics{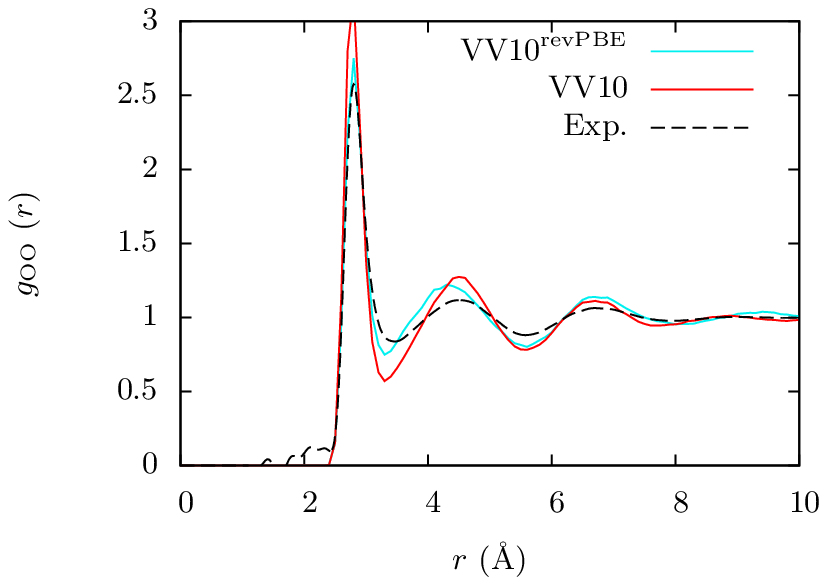}}
\subfigure[\ Functionals based on vdW-DF]{\includegraphics{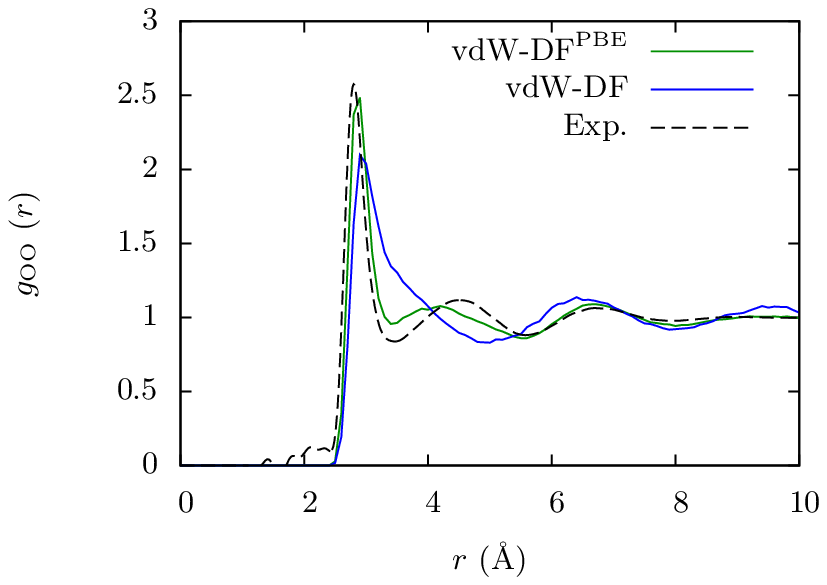}}
\caption{Comparison of the O--O RDFs from AIMD simulations (1.00~g/cm$^3$, $\sim$300~K) with experimental data at ambient conditions~[\onlinecite{waterRDF2013}].}
\label{appfig:rdf_exp}
\end{figure*}

Fig.~\ref{appfig:rdf_exp} shows the comparison of the RDFs from AIMD simulations with experiment for four different xc functionals. All simulations are performed at the same density and temperature, although with differing system sizes (see Table~\ref{apptable:all_runs}). The experimental data from x-ray diffraction measurements~\cite{waterRDF2013} is for ambient conditions.

\begin{figure*}
\subfigure[\ Position of maxima/minima]{\includegraphics{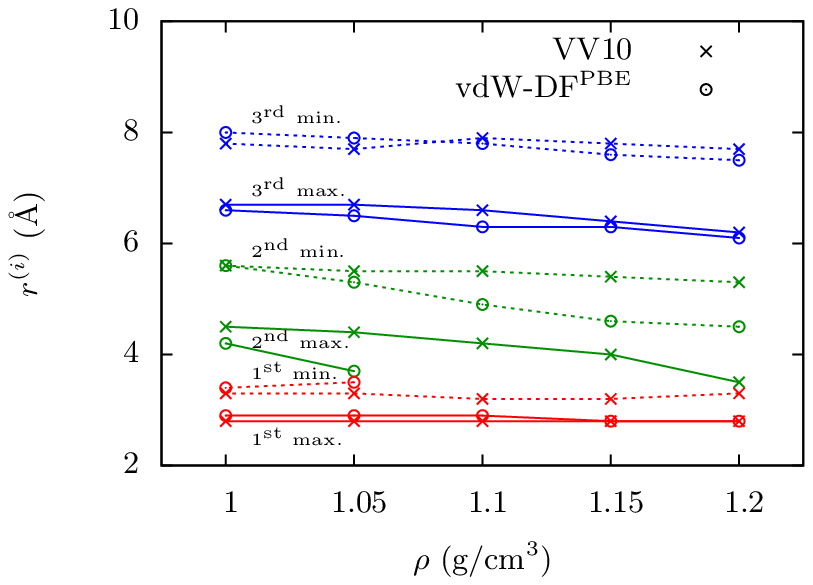}}
\subfigure[\ Height of maxima/minima]{\includegraphics{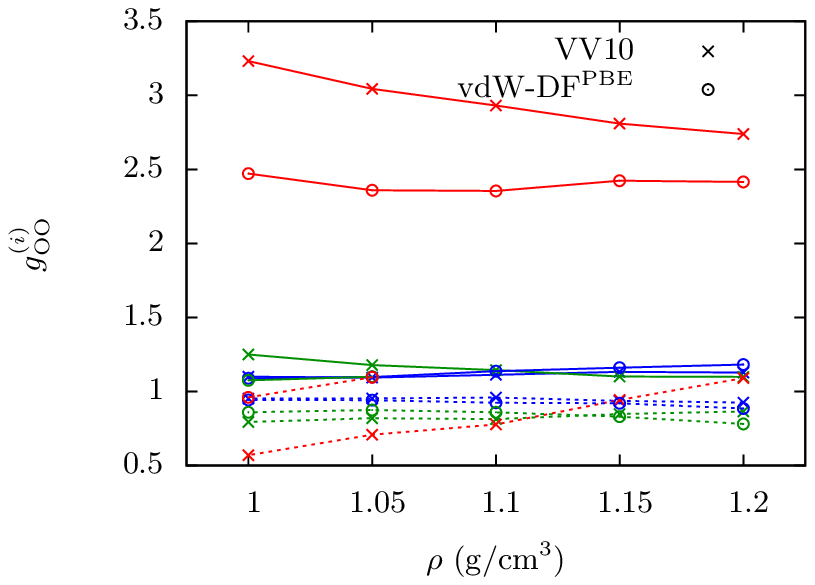}}
\caption{Variation with density in the position and height of the first three maxima and minima for the RDFs shown in Fig.~\ref{fig:rdf_trends}.}
\label{appfig:rdf_trends}
\end{figure*}

Finally, Fig.~\ref{appfig:rdf_trends} shows how the radial position and height of the first three maxima and minima in the RDF change with density at $\sim$300~K for VV10 and vdW-DF$^\mathrm{PBE}$.

\section{H-bond definition}
\label{appendix-HB}

We describe here our criterion for the existence of a H bond between two molecules, used in the RDF decomposition of Fig.~\ref{fig:rdf_decomp}, as well as previously in Wang {\em et al.}~\cite{water_emiliomarivi}.

In the standard geometrical definition, a H bond exists if two conditions are satisfied: (i) the intermolecular distance $r_\mathrm{OO} < r_\mathrm{OO}^\mathrm{cut}$, where $r_\mathrm{OO}^\mathrm{cut}$ is usually chosen as the position of the first minimum in the O--O RDF ($\sim$3.5~\AA), and (ii) the angle $\widehat{\mathrm{O}_a \mathrm{O}_d \mathrm{H}_d} < \theta_\mathrm{OOH}^\mathrm{cut}$, where $a,d$ indicate the acceptor or donor character of the molecules participating in the bond, and $\theta_\mathrm{OOH}^\mathrm{cut} \sim 30^\circ$ is a cutoff angle. However, this definition ignores the electronic distribution of the acceptor molecule, which plays an essential role in H bonds~\cite{Fernandez-Serra2006,Kumar2007}. 

We consider a donor H atom and an acceptor lone pair (L), and we define that a H bond exists if their distance $r_\mathrm{HL} < r_\mathrm{HL}^\mathrm{cut}=2.0$~\AA. The lone pair centers can be obtained from their maximally localized Wannier orbitals~\cite{Sharma2005}, but the cost of this calculation at every time step would be prohibitive. Instead, assuming that the two bonding orbitals are in the OH directions, we use the orthogonality of the four $sp^3$ hybrid orbitals to determine the angle $\theta_\mathrm{LOL}$ between the lone pair directions:
\begin{equation}
\left [ 1-\tan^2(\theta_\mathrm{HOH}/2) \right ] 
\left [ 1-\tan^2(\theta_\mathrm{LOL}/2) \right ] = 1.
\end{equation}
Then, the positions of the two lone pair centers are
\begin{equation}
\mathbf{r}_\mathrm{OL}=r_\mathrm{OL} \left[ - \mathbf{u}_b\cos(\theta_\mathrm{LOL}/2) \pm \mathbf{u}_n\sin(\theta_\mathrm{LOL}/2) \right]
\end{equation}
where $\mathbf{u}_b$ and $\mathbf{u}_n$ are unit vectors along the HOH bisector and normal to the molecular plane, respectively, and $r_\mathrm{OL}$ is the distance of the lone pair centers to the oxygen atom. We use the value of the TIP5P model~\cite{TIP5P} ($r_\mathrm{OL} = 0.7$~\AA) which is very close to that of the ST2 model~\cite{ST2} (0.8~\AA). Although this is about two times longer than the distance of Wannier orbitals, we have found that the H-bond definition is very insensitive to $r_\mathrm{OL}$ and $r_\mathrm{HL}^\mathrm{cut}$, provided that $r_\mathrm{OH} + r_\mathrm{OL} + r_\mathrm{HL}^\mathrm{cut} \simeq 3.6$~\AA. We have also checked that this definition produces very similar results to those of an electronic-based criterion, such as the Mulliken overlap~\cite{Fernandez-Serra2006}.

\section{Static and dynamic pressure differences}
\label{appendix-P}

\begin{table*}
\caption{Pressure differences between xc functionals (1.00 g/cm$^3$, $\sim$300 K). For each pair of functionals we list the total difference $\Delta P$, and its static and dynamic components ($\Delta P_s$ and $\Delta P_d$, respectively). PBE and revPBE data from Ref.~[\onlinecite{water_emiliomarivi}] (basis set and temperature corrections applied). All values in kbar.}
\label{apptable:grid}
\begin{ruledtabular}
{\scriptsize \begin{tabular*}{\textwidth}{l r r r r r r r r r r r r r r r r r r r}
& \multicolumn{3}{c}{PBE}   & \multicolumn{3}{c}{revPBE} & \multicolumn{3}{c}{PBE$^\mathrm{PW86R}$} & \multicolumn{3}{c}{vdW-DF}  & \multicolumn{3}{c}{vdW-DF$^\mathrm{PBE}$} & \multicolumn{3}{c}{VV10} \\ \cline{2-4} \cline{5-7} \cline{8-10} \cline{11-13} \cline{14-16} \cline{17-19}
& $\Delta P$ & $\Delta P_\mathrm{s}$ & $\Delta P_\mathrm{d}$ & $\Delta P$ & $\Delta P_\mathrm{s}$ & $\Delta P_\mathrm{d}$ & $\Delta P$ & $\Delta P_\mathrm{s}$ & $\Delta P_\mathrm{d}$ & $\Delta P$ & $\Delta P_\mathrm{s}$ & $\Delta P_\mathrm{d}$ & $\Delta P$ & $\Delta P_\mathrm{s}$ & $\Delta P_\mathrm{d}$ & $\Delta P$ & $\Delta P_\mathrm{s}$ & $\Delta P_\mathrm{d}$ \\
\hline
PBE                   & \multicolumn{3}{c}{-} \\
revPBE                &     7.8 &    7.3 &    0.5 & \multicolumn{3}{c}{-} \\
PBE$^\mathrm{PW86R}$  &  $-$0.3 &    1.1 & $-$1.4 &  $-$7.5 &  $-$6.2 & $-$1.3 & \multicolumn{3}{c}{-} \\
vdW-DF                &  $-$2.3 &    1.4 & $-$3.7 & $-$10.1 &  $-$5.9 & $-$4.2 &  $-$2.6 &    0.3 & $-$2.9    & \multicolumn{3}{c}{-} \\
vdW-DF$^\mathrm{PBE}$ & $-$11.7 & $-$6.0 & $-$5.7 & $-$19.5 & $-$13.3 & $-$6.2 & $-$11.9 & $-$7.1 & $-$4.8    & $-$9.4 & $-$7.4 & $-$2.0   & \multicolumn{3}{c}{-} \\
VV10                  &  $-$7.5 & $-$6.2 & $-$1.3 & $-$15.3 & $-$13.5 & $-$1.8 &  $-$7.8 & $-$7.3 & $-$0.5    & $-$5.2 & $-$7.6 &    2.4   &  4.2 & $-$0.2 & 4.4           & \multicolumn{3}{c}{-} \\
VV10$^\mathrm{PBE}$   &  $-$1.7 &    0.3 & $-$1.4 &  $-$9.5 &  $-$7.0 & $-$2.5 &  $-$2.0 & $-$0.8 & $-$1.2    &    0.6 & $-$1.1 & $-$0.5   & 10.0 &    6.3 & 3.7           & 5.8 & 6.4 & $-$0.6 \\
\end{tabular*}}
\end{ruledtabular}
\end{table*}

In order to investigate the variation in average pressure in the AIMD simulations due to the xc functional, we have calculated `static' and `dynamic' contributions to the total pressure difference between pairs of functionals, listed in Table~\ref{apptable:grid}. We make use of three GGA functionals and four vdW functionals (two based on vdW-DF, and two based on VV10).

The total difference $\Delta P$ is calculated using the average pressures $P_\mathrm{corr}$ obtained from the AIMD simulations (see Table~\ref{apptable:all_runs}). Instead, the static difference $\Delta P_s$ is obtained from the average pressures calculated using the {\em same} 200 snapshots of liquid water for both functionals. These include both low- and high-density configurations of the liquid. We find that there is an almost constant difference between functionals when calculating $P$ from the same snapshot. Finally, the dynamic difference $\Delta P_d$ is taken as the discrepancy between $\Delta P_s$ and $\Delta P$. A large value of $\Delta P_d$, therefore, indicates that the two functionals are exploring significantly different configurations during the AIMD simulation.

From the results presented in the table, we find that the largest values of $\Delta P_d$ are for differences between vdW-DF and GGA functionals (4.7~kbar RMS), followed by differences between vdW-DF and VV10 functionals (3.1~kbar RMS). Instead, differences between VV10 and GGA functionals are much smaller (1.6~kbar RMS), as well as those between different GGA functionals (1.1~kbar RMS). Therefore, this suggests that vdW-DF significantly alters the sampling of configuration space compared to GGA, while VV10 has only a small effect in this respect.


\begin{thebibliography}{68}%
\makeatletter
\providecommand \@ifxundefined [1]{%
 \@ifx{#1\undefined}
}%
\providecommand \@ifnum [1]{%
 \ifnum #1\expandafter \@firstoftwo
 \else \expandafter \@secondoftwo
 \fi
}%
\providecommand \@ifx [1]{%
 \ifx #1\expandafter \@firstoftwo
 \else \expandafter \@secondoftwo
 \fi
}%
\providecommand \natexlab [1]{#1}%
\providecommand \enquote  [1]{``#1''}%
\providecommand \bibnamefont  [1]{#1}%
\providecommand \bibfnamefont [1]{#1}%
\providecommand \citenamefont [1]{#1}%
\providecommand \href@noop [0]{\@secondoftwo}%
\providecommand \href [0]{\begingroup \@sanitize@url \@href}%
\providecommand \@href[1]{\@@startlink{#1}\@@href}%
\providecommand \@@href[1]{\endgroup#1\@@endlink}%
\providecommand \@sanitize@url [0]{\catcode `\\12\catcode `\$12\catcode
  `\&12\catcode `\#12\catcode `\^12\catcode `\_12\catcode `\%12\relax}%
\providecommand \@@startlink[1]{}%
\providecommand \@@endlink[0]{}%
\providecommand \url  [0]{\begingroup\@sanitize@url \@url }%
\providecommand \@url [1]{\endgroup\@href {#1}{\urlprefix }}%
\providecommand \urlprefix  [0]{URL }%
\providecommand \Eprint [0]{\href }%
\providecommand \doibase [0]{http://dx.doi.org/}%
\providecommand \selectlanguage [0]{\@gobble}%
\providecommand \bibinfo  [0]{\@secondoftwo}%
\providecommand \bibfield  [0]{\@secondoftwo}%
\providecommand \translation [1]{[#1]}%
\providecommand \BibitemOpen [0]{}%
\providecommand \bibitemStop [0]{}%
\providecommand \bibitemNoStop [0]{.\EOS\space}%
\providecommand \EOS [0]{\spacefactor3000\relax}%
\providecommand \BibitemShut  [1]{\csname bibitem#1\endcsname}%
\let\auto@bib@innerbib\@empty
\bibitem [{\citenamefont {Guillot}(2002)}]{Guillot2002}%
  \BibitemOpen
  \bibfield  {author} {\bibinfo {author} {\bibfnamefont {B.}~\bibnamefont
  {Guillot}},\ }\href@noop {} {\bibfield  {journal} {\bibinfo  {journal} {J.
  Mol. Liq.}\ }\textbf {\bibinfo {volume} {101}},\ \bibinfo {pages} {219}
  (\bibinfo {year} {2002})}\BibitemShut {NoStop}%
\bibitem [{\citenamefont {Paschek}(2004)}]{Paschek2004}%
  \BibitemOpen
  \bibfield  {author} {\bibinfo {author} {\bibfnamefont {D.}~\bibnamefont
  {Paschek}},\ }\href@noop {} {\bibfield  {journal} {\bibinfo  {journal} {J.
  Chem. Phys.}\ }\textbf {\bibinfo {volume} {120}},\ \bibinfo {pages} {6674}
  (\bibinfo {year} {2004})}\BibitemShut {NoStop}%
\bibitem [{\citenamefont {Abascal}\ and\ \citenamefont
  {Vega}(2005)}]{Abascal2005}%
  \BibitemOpen
  \bibfield  {author} {\bibinfo {author} {\bibfnamefont {J.~L.~F.}\
  \bibnamefont {Abascal}}\ and\ \bibinfo {author} {\bibfnamefont
  {C.}~\bibnamefont {Vega}},\ }\href {\doibase 10.1063/1.2121687} {\bibfield
  {journal} {\bibinfo  {journal} {J. Chem. Phys.}\ }\textbf {\bibinfo {volume}
  {123}},\ \bibinfo {eid} {234505} (\bibinfo {year} {2005})}\BibitemShut
  {NoStop}%
\bibitem [{\citenamefont {Pi}\ \emph {et~al.}(2009)\citenamefont {Pi},
  \citenamefont {Aragones}, \citenamefont {Vega}, \citenamefont {Noya},
  \citenamefont {Abascal}, \citenamefont {Gonzalez},\ and\ \citenamefont
  {McBride}}]{Vega2009}%
  \BibitemOpen
  \bibfield  {author} {\bibinfo {author} {\bibfnamefont {H.~L.}\ \bibnamefont
  {Pi}}, \bibinfo {author} {\bibfnamefont {J.~L.}\ \bibnamefont {Aragones}},
  \bibinfo {author} {\bibfnamefont {C.}~\bibnamefont {Vega}}, \bibinfo {author}
  {\bibfnamefont {E.~G.}\ \bibnamefont {Noya}}, \bibinfo {author}
  {\bibfnamefont {J.~L.~F.}\ \bibnamefont {Abascal}}, \bibinfo {author}
  {\bibfnamefont {M.~A.}\ \bibnamefont {Gonzalez}}, \ and\ \bibinfo {author}
  {\bibfnamefont {C.}~\bibnamefont {McBride}},\ }\href@noop {} {\bibfield
  {journal} {\bibinfo  {journal} {Mol. Phys.}\ }\textbf {\bibinfo {volume}
  {107}},\ \bibinfo {pages} {365} (\bibinfo {year} {2009})}\BibitemShut
  {NoStop}%
\bibitem [{\citenamefont {Sciortino}\ \emph {et~al.}(1997)\citenamefont
  {Sciortino}, \citenamefont {Poole}, \citenamefont {Essmann},\ and\
  \citenamefont {Stanley}}]{Sciortino1997}%
  \BibitemOpen
  \bibfield  {author} {\bibinfo {author} {\bibfnamefont {F.}~\bibnamefont
  {Sciortino}}, \bibinfo {author} {\bibfnamefont {P.~H.}\ \bibnamefont
  {Poole}}, \bibinfo {author} {\bibfnamefont {U.}~\bibnamefont {Essmann}}, \
  and\ \bibinfo {author} {\bibfnamefont {H.~E.}\ \bibnamefont {Stanley}},\
  }\href {\doibase 10.1103/PhysRevE.55.727} {\bibfield  {journal} {\bibinfo
  {journal} {Phys. Rev. E}\ }\textbf {\bibinfo {volume} {55}},\ \bibinfo
  {pages} {727} (\bibinfo {year} {1997})}\BibitemShut {NoStop}%
\bibitem [{\citenamefont {Moore}\ and\ \citenamefont
  {Molinero}(2011)}]{Moore2011}%
  \BibitemOpen
  \bibfield  {author} {\bibinfo {author} {\bibfnamefont {E.~B.}\ \bibnamefont
  {Moore}}\ and\ \bibinfo {author} {\bibfnamefont {V.}~\bibnamefont
  {Molinero}},\ }\href {http://dx.doi.org/10.1038/nature10586
  http://www.nature.com/nature/journal/v479/n7374/abs/nature10586.html\#supplementary-information}
  {\bibfield  {journal} {\bibinfo  {journal} {Nature}\ }\textbf {\bibinfo
  {volume} {479}},\ \bibinfo {pages} {506} (\bibinfo {year}
  {2011})}\BibitemShut {NoStop}%
\bibitem [{\citenamefont {Limmer}\ and\ \citenamefont
  {Chandler}(2012)}]{Chandler2012}%
  \BibitemOpen
  \bibfield  {author} {\bibinfo {author} {\bibfnamefont {D.~T.}\ \bibnamefont
  {Limmer}}\ and\ \bibinfo {author} {\bibfnamefont {D.}~\bibnamefont
  {Chandler}},\ }\href {\doibase 10.1063/1.4737907} {\bibfield  {journal}
  {\bibinfo  {journal} {J. Chem. Phys.}\ }\textbf {\bibinfo {volume} {137}},\
  \bibinfo {eid} {044509} (\bibinfo {year} {2012})}\BibitemShut {NoStop}%
\bibitem [{\citenamefont {Mallamace}, \citenamefont {Corsaro},\ and\
  \citenamefont {Stanley}(2013)}]{MallamacePNAS2013}%
  \BibitemOpen
  \bibfield  {author} {\bibinfo {author} {\bibfnamefont {F.}~\bibnamefont
  {Mallamace}}, \bibinfo {author} {\bibfnamefont {C.}~\bibnamefont {Corsaro}},
  \ and\ \bibinfo {author} {\bibfnamefont {H.~E.}\ \bibnamefont {Stanley}},\
  }\href {\doibase 10.1073/pnas.1221805110} {\bibfield  {journal} {\bibinfo
  {journal} {Proc. Natl. Acad. Sci. USA}\ }\textbf {\bibinfo {volume} {110}},\
  \bibinfo {pages} {4899} (\bibinfo {year} {2013})}\BibitemShut {NoStop}%
\bibitem [{\citenamefont {Huang}\ \emph {et~al.}(2009)\citenamefont {Huang},
  \citenamefont {Wikfeldt}, \citenamefont {Tokushima}, \citenamefont
  {Nordlund}, \citenamefont {Harada}, \citenamefont {Bergmann}, \citenamefont
  {Niebuhr}, \citenamefont {Weiss}, \citenamefont {Horikawa}, \citenamefont
  {Leetmaa}, \citenamefont {Ljungberg}, \citenamefont {Takahashi},
  \citenamefont {Lenz}, \citenamefont {Ojam\"{a}e}, \citenamefont {Lyubartsev},
  \citenamefont {Shin}, \citenamefont {Pettersson},\ and\ \citenamefont
  {Nilsson}}]{Huang2009}%
  \BibitemOpen
  \bibfield  {author} {\bibinfo {author} {\bibfnamefont {C.}~\bibnamefont
  {Huang}}, \bibinfo {author} {\bibfnamefont {K.~T.}\ \bibnamefont {Wikfeldt}},
  \bibinfo {author} {\bibfnamefont {T.}~\bibnamefont {Tokushima}}, \bibinfo
  {author} {\bibfnamefont {D.}~\bibnamefont {Nordlund}}, \bibinfo {author}
  {\bibfnamefont {Y.}~\bibnamefont {Harada}}, \bibinfo {author} {\bibfnamefont
  {U.}~\bibnamefont {Bergmann}}, \bibinfo {author} {\bibfnamefont
  {M.}~\bibnamefont {Niebuhr}}, \bibinfo {author} {\bibfnamefont {T.~M.}\
  \bibnamefont {Weiss}}, \bibinfo {author} {\bibfnamefont {Y.}~\bibnamefont
  {Horikawa}}, \bibinfo {author} {\bibfnamefont {M.}~\bibnamefont {Leetmaa}},
  \bibinfo {author} {\bibfnamefont {M.~P.}\ \bibnamefont {Ljungberg}}, \bibinfo
  {author} {\bibfnamefont {O.}~\bibnamefont {Takahashi}}, \bibinfo {author}
  {\bibfnamefont {A.}~\bibnamefont {Lenz}}, \bibinfo {author} {\bibfnamefont
  {L.}~\bibnamefont {Ojam\"{a}e}}, \bibinfo {author} {\bibfnamefont {A.~P.}\
  \bibnamefont {Lyubartsev}}, \bibinfo {author} {\bibfnamefont
  {S.}~\bibnamefont {Shin}}, \bibinfo {author} {\bibfnamefont {L.~G.~M.}\
  \bibnamefont {Pettersson}}, \ and\ \bibinfo {author} {\bibfnamefont
  {A.}~\bibnamefont {Nilsson}},\ }\href@noop {} {\bibfield  {journal} {\bibinfo
   {journal} {Proc. Natl. Acad. Sci. USA}\ }\textbf {\bibinfo {volume} {106}},\
  \bibinfo {pages} {15214} (\bibinfo {year} {2009})}\BibitemShut {NoStop}%
\bibitem [{\citenamefont {Kumar}\ and\ \citenamefont
  {Stanley}(2011)}]{Kumar2011}%
  \BibitemOpen
  \bibfield  {author} {\bibinfo {author} {\bibfnamefont {P.}~\bibnamefont
  {Kumar}}\ and\ \bibinfo {author} {\bibfnamefont {H.~E.}\ \bibnamefont
  {Stanley}},\ }\href {\doibase 10.1021/jp2051867} {\bibfield  {journal}
  {\bibinfo  {journal} {J. Phys. Chem. B}\ }\textbf {\bibinfo {volume} {115}},\
  \bibinfo {pages} {14269} (\bibinfo {year} {2011})}\BibitemShut {NoStop}%
\bibitem [{\citenamefont {Sciortino}, \citenamefont {Saika-Voivod},\ and\
  \citenamefont {Poole}(2011)}]{Sciortino2011}%
  \BibitemOpen
  \bibfield  {author} {\bibinfo {author} {\bibfnamefont {F.}~\bibnamefont
  {Sciortino}}, \bibinfo {author} {\bibfnamefont {I.}~\bibnamefont
  {Saika-Voivod}}, \ and\ \bibinfo {author} {\bibfnamefont {P.~H.}\
  \bibnamefont {Poole}},\ }\href@noop {} {\bibfield  {journal} {\bibinfo
  {journal} {Phys. Chem. Chem. Phys.}\ }\textbf {\bibinfo {volume} {13}},\
  \bibinfo {pages} {19759} (\bibinfo {year} {2011})}\BibitemShut {NoStop}%
\bibitem [{\citenamefont {Abascal}\ and\ \citenamefont
  {Vega}(2011)}]{Abascal2011}%
  \BibitemOpen
  \bibfield  {author} {\bibinfo {author} {\bibfnamefont {J.~L.~F.}\
  \bibnamefont {Abascal}}\ and\ \bibinfo {author} {\bibfnamefont
  {C.}~\bibnamefont {Vega}},\ }\href {\doibase 10.1063/1.3585676} {\bibfield
  {journal} {\bibinfo  {journal} {J. Chem. Phys.}\ }\textbf {\bibinfo {volume}
  {134}},\ \bibinfo {eid} {186101} (\bibinfo {year} {2011})}\BibitemShut
  {NoStop}%
\bibitem [{\citenamefont {Paschek}(2005)}]{Paschek2005}%
  \BibitemOpen
  \bibfield  {author} {\bibinfo {author} {\bibfnamefont {D.}~\bibnamefont
  {Paschek}},\ }\href {\doibase 10.1103/PhysRevLett.94.217802} {\bibfield
  {journal} {\bibinfo  {journal} {Phys. Rev. Lett.}\ }\textbf {\bibinfo
  {volume} {94}},\ \bibinfo {pages} {217802} (\bibinfo {year}
  {2005})}\BibitemShut {NoStop}%
\bibitem [{\citenamefont {Corradini}, \citenamefont {Rovere},\ and\
  \citenamefont {Gallo}(2010)}]{Gallo2010}%
  \BibitemOpen
  \bibfield  {author} {\bibinfo {author} {\bibfnamefont {D.}~\bibnamefont
  {Corradini}}, \bibinfo {author} {\bibfnamefont {M.}~\bibnamefont {Rovere}}, \
  and\ \bibinfo {author} {\bibfnamefont {P.}~\bibnamefont {Gallo}},\ }\href
  {\doibase 10.1063/1.3376776} {\bibfield  {journal} {\bibinfo  {journal} {J.
  Chem. Phys.}\ }\textbf {\bibinfo {volume} {132}},\ \bibinfo {eid} {134508}
  (\bibinfo {year} {2010})}\BibitemShut {NoStop}%
\bibitem [{\citenamefont {Abascal}\ and\ \citenamefont
  {Vega}(2010)}]{Abascal2010}%
  \BibitemOpen
  \bibfield  {author} {\bibinfo {author} {\bibfnamefont {J.~L.~F.}\
  \bibnamefont {Abascal}}\ and\ \bibinfo {author} {\bibfnamefont
  {C.}~\bibnamefont {Vega}},\ }\href {\doibase 10.1063/1.3506860} {\bibfield
  {journal} {\bibinfo  {journal} {J. Chem. Phys.}\ }\textbf {\bibinfo {volume}
  {133}},\ \bibinfo {eid} {234502} (\bibinfo {year} {2010})}\BibitemShut
  {NoStop}%
\bibitem [{\citenamefont {Pamuk}\ \emph {et~al.}(2012)\citenamefont {Pamuk},
  \citenamefont {Soler}, \citenamefont {Ram\'{i}rez}, \citenamefont {Herrero},
  \citenamefont {Stephens}, \citenamefont {Allen},\ and\ \citenamefont
  {Fern\'{a}ndez-Serra}}]{Pamuk2012}%
  \BibitemOpen
  \bibfield  {author} {\bibinfo {author} {\bibfnamefont {B.}~\bibnamefont
  {Pamuk}}, \bibinfo {author} {\bibfnamefont {J.~M.}\ \bibnamefont {Soler}},
  \bibinfo {author} {\bibfnamefont {R.}~\bibnamefont {Ram\'{i}rez}}, \bibinfo
  {author} {\bibfnamefont {C.~P.}\ \bibnamefont {Herrero}}, \bibinfo {author}
  {\bibfnamefont {P.~W.}\ \bibnamefont {Stephens}}, \bibinfo {author}
  {\bibfnamefont {P.~B.}\ \bibnamefont {Allen}}, \ and\ \bibinfo {author}
  {\bibfnamefont {M.-V.}\ \bibnamefont {Fern\'{a}ndez-Serra}},\ }\href
  {\doibase 10.1103/PhysRevLett.108.193003} {\bibfield  {journal} {\bibinfo
  {journal} {Phys. Rev. Lett.}\ }\textbf {\bibinfo {volume} {108}},\ \bibinfo
  {pages} {193003} (\bibinfo {year} {2012})}\BibitemShut {NoStop}%
\bibitem [{\citenamefont {Grossman}\ \emph {et~al.}(2004)\citenamefont
  {Grossman}, \citenamefont {Schwegler}, \citenamefont {Draeger}, \citenamefont
  {Gygi},\ and\ \citenamefont {Galli}}]{Grossman2004}%
  \BibitemOpen
  \bibfield  {author} {\bibinfo {author} {\bibfnamefont {J.~C.}\ \bibnamefont
  {Grossman}}, \bibinfo {author} {\bibfnamefont {E.}~\bibnamefont {Schwegler}},
  \bibinfo {author} {\bibfnamefont {E.~W.}\ \bibnamefont {Draeger}}, \bibinfo
  {author} {\bibfnamefont {F.}~\bibnamefont {Gygi}}, \ and\ \bibinfo {author}
  {\bibfnamefont {G.}~\bibnamefont {Galli}},\ }\href@noop {} {\bibfield
  {journal} {\bibinfo  {journal} {J. Chem. Phys.}\ }\textbf {\bibinfo {volume}
  {120}},\ \bibinfo {pages} {300} (\bibinfo {year} {2004})}\BibitemShut
  {NoStop}%
\bibitem [{\citenamefont {Fern\'{a}ndez-Serra}\ and\ \citenamefont
  {Artacho}(2004)}]{Fernandez-Serra2004}%
  \BibitemOpen
  \bibfield  {author} {\bibinfo {author} {\bibfnamefont {M.-V.}\ \bibnamefont
  {Fern\'{a}ndez-Serra}}\ and\ \bibinfo {author} {\bibfnamefont
  {E.}~\bibnamefont {Artacho}},\ }\href@noop {} {\bibfield  {journal} {\bibinfo
   {journal} {J. Chem. Phys.}\ }\textbf {\bibinfo {volume} {121}},\ \bibinfo
  {pages} {11136} (\bibinfo {year} {2004})}\BibitemShut {NoStop}%
\bibitem [{\citenamefont {Kuo}\ \emph {et~al.}(2004)\citenamefont {Kuo},
  \citenamefont {Mundy}, \citenamefont {McGrath}, \citenamefont {Siepmann},
  \citenamefont {VandeVondele}, \citenamefont {Sprik}, \citenamefont {Hutter},
  \citenamefont {Chen}, \citenamefont {Klein}, \citenamefont {Mohamed},
  \citenamefont {Krack},\ and\ \citenamefont {Parrinello}}]{Kuo2004}%
  \BibitemOpen
  \bibfield  {author} {\bibinfo {author} {\bibfnamefont {I.-F.~W.}\
  \bibnamefont {Kuo}}, \bibinfo {author} {\bibfnamefont {C.~J.}\ \bibnamefont
  {Mundy}}, \bibinfo {author} {\bibfnamefont {M.~J.}\ \bibnamefont {McGrath}},
  \bibinfo {author} {\bibfnamefont {J.~I.}\ \bibnamefont {Siepmann}}, \bibinfo
  {author} {\bibfnamefont {J.}~\bibnamefont {VandeVondele}}, \bibinfo {author}
  {\bibfnamefont {M.}~\bibnamefont {Sprik}}, \bibinfo {author} {\bibfnamefont
  {J.}~\bibnamefont {Hutter}}, \bibinfo {author} {\bibfnamefont
  {B.}~\bibnamefont {Chen}}, \bibinfo {author} {\bibfnamefont {M.~L.}\
  \bibnamefont {Klein}}, \bibinfo {author} {\bibfnamefont {F.}~\bibnamefont
  {Mohamed}}, \bibinfo {author} {\bibfnamefont {M.}~\bibnamefont {Krack}}, \
  and\ \bibinfo {author} {\bibfnamefont {M.}~\bibnamefont {Parrinello}},\
  }\href@noop {} {\bibfield  {journal} {\bibinfo  {journal} {J. Phys. Chem. B}\
  }\textbf {\bibinfo {volume} {108}},\ \bibinfo {pages} {12990} (\bibinfo
  {year} {2004})}\BibitemShut {NoStop}%
\bibitem [{\citenamefont {Sit}\ and\ \citenamefont {Marzari}(2005)}]{Sit2005}%
  \BibitemOpen
  \bibfield  {author} {\bibinfo {author} {\bibfnamefont {P.~H.-L.}\
  \bibnamefont {Sit}}\ and\ \bibinfo {author} {\bibfnamefont {N.}~\bibnamefont
  {Marzari}},\ }\href@noop {} {\bibfield  {journal} {\bibinfo  {journal} {J.
  Chem. Phys.}\ }\textbf {\bibinfo {volume} {122}},\ \bibinfo {pages} {204510}
  (\bibinfo {year} {2005})}\BibitemShut {NoStop}%
\bibitem [{\citenamefont {Schmidt}\ \emph {et~al.}(2009)\citenamefont
  {Schmidt}, \citenamefont {VandeVondele}, \citenamefont {Kuo}, \citenamefont
  {Sebastiani}, \citenamefont {Siepmann}, \citenamefont {Hutter},\ and\
  \citenamefont {Mundy}}]{Schmidt2009}%
  \BibitemOpen
  \bibfield  {author} {\bibinfo {author} {\bibfnamefont {J.}~\bibnamefont
  {Schmidt}}, \bibinfo {author} {\bibfnamefont {J.}~\bibnamefont
  {VandeVondele}}, \bibinfo {author} {\bibfnamefont {I.-F.~W.}\ \bibnamefont
  {Kuo}}, \bibinfo {author} {\bibfnamefont {D.}~\bibnamefont {Sebastiani}},
  \bibinfo {author} {\bibfnamefont {J.~I.}\ \bibnamefont {Siepmann}}, \bibinfo
  {author} {\bibfnamefont {J.}~\bibnamefont {Hutter}}, \ and\ \bibinfo {author}
  {\bibfnamefont {C.~J.}\ \bibnamefont {Mundy}},\ }\href@noop {} {\bibfield
  {journal} {\bibinfo  {journal} {J. Phys. Chem. B}\ }\textbf {\bibinfo
  {volume} {113}},\ \bibinfo {pages} {11959} (\bibinfo {year}
  {2009})}\BibitemShut {NoStop}%
\bibitem [{\citenamefont {Dion}\ \emph {et~al.}(2004)\citenamefont {Dion},
  \citenamefont {Rydberg}, \citenamefont {Schr\"{o}der}, \citenamefont
  {Langreth},\ and\ \citenamefont {Lundqvist}}]{vdW-DF}%
  \BibitemOpen
  \bibfield  {author} {\bibinfo {author} {\bibfnamefont {M.}~\bibnamefont
  {Dion}}, \bibinfo {author} {\bibfnamefont {H.}~\bibnamefont {Rydberg}},
  \bibinfo {author} {\bibfnamefont {E.}~\bibnamefont {Schr\"{o}der}}, \bibinfo
  {author} {\bibfnamefont {D.~C.}\ \bibnamefont {Langreth}}, \ and\ \bibinfo
  {author} {\bibfnamefont {B.~I.}\ \bibnamefont {Lundqvist}},\ }\href@noop {}
  {\bibfield  {journal} {\bibinfo  {journal} {Phys. Rev. Lett.}\ }\textbf
  {\bibinfo {volume} {92}},\ \bibinfo {pages} {246401} (\bibinfo {year}
  {2004})}\BibitemShut {NoStop}%
\bibitem [{\citenamefont {Klimes}, \citenamefont {Bowler},\ and\ \citenamefont
  {Michaelides}(2010)}]{Klimes2010}%
  \BibitemOpen
  \bibfield  {author} {\bibinfo {author} {\bibfnamefont {J.}~\bibnamefont
  {Klimes}}, \bibinfo {author} {\bibfnamefont {D.~R.}\ \bibnamefont {Bowler}},
  \ and\ \bibinfo {author} {\bibfnamefont {A.}~\bibnamefont {Michaelides}},\
  }\href@noop {} {\bibfield  {journal} {\bibinfo  {journal} {J. Phys.: Condens.
  Matter}\ }\textbf {\bibinfo {volume} {22}},\ \bibinfo {pages} {022201}
  (\bibinfo {year} {2010})}\BibitemShut {NoStop}%
\bibitem [{\citenamefont {Lee}\ \emph {et~al.}(2010)\citenamefont {Lee},
  \citenamefont {Murray}, \citenamefont {Kong}, \citenamefont {Lundqvist},\
  and\ \citenamefont {Langreth}}]{Lee2010}%
  \BibitemOpen
  \bibfield  {author} {\bibinfo {author} {\bibfnamefont {K.}~\bibnamefont
  {Lee}}, \bibinfo {author} {\bibfnamefont {E.~D.}\ \bibnamefont {Murray}},
  \bibinfo {author} {\bibfnamefont {L.}~\bibnamefont {Kong}}, \bibinfo {author}
  {\bibfnamefont {B.~I.}\ \bibnamefont {Lundqvist}}, \ and\ \bibinfo {author}
  {\bibfnamefont {D.~C.}\ \bibnamefont {Langreth}},\ }\href@noop {} {\bibfield
  {journal} {\bibinfo  {journal} {Phys. Rev. B}\ }\textbf {\bibinfo {volume}
  {82}},\ \bibinfo {pages} {081101(R)} (\bibinfo {year} {2010})}\BibitemShut
  {NoStop}%
\bibitem [{\citenamefont {Vydrov}\ and\ \citenamefont
  {Van~Voorhis}(2010)}]{VV10}%
  \BibitemOpen
  \bibfield  {author} {\bibinfo {author} {\bibfnamefont {O.~A.}\ \bibnamefont
  {Vydrov}}\ and\ \bibinfo {author} {\bibfnamefont {T.}~\bibnamefont
  {Van~Voorhis}},\ }\href@noop {} {\bibfield  {journal} {\bibinfo  {journal}
  {J. Chem. Phys.}\ }\textbf {\bibinfo {volume} {133}},\ \bibinfo {pages}
  {244103} (\bibinfo {year} {2010})}\BibitemShut {NoStop}%
\bibitem [{\citenamefont {Lin}\ \emph {et~al.}(2009)\citenamefont {Lin},
  \citenamefont {Seitsonen}, \citenamefont {Coutinho-Neto}, \citenamefont
  {Tavernelli},\ and\ \citenamefont {Rothlisberger}}]{Lin2009}%
  \BibitemOpen
  \bibfield  {author} {\bibinfo {author} {\bibfnamefont {I.-C.}\ \bibnamefont
  {Lin}}, \bibinfo {author} {\bibfnamefont {A.~P.}\ \bibnamefont {Seitsonen}},
  \bibinfo {author} {\bibfnamefont {M.~D.}\ \bibnamefont {Coutinho-Neto}},
  \bibinfo {author} {\bibfnamefont {I.}~\bibnamefont {Tavernelli}}, \ and\
  \bibinfo {author} {\bibfnamefont {U.}~\bibnamefont {Rothlisberger}},\ }\href
  {\doibase 10.1021/jp806376e} {\bibfield  {journal} {\bibinfo  {journal} {J.
  Phys. Chem. B}\ }\textbf {\bibinfo {volume} {113}},\ \bibinfo {pages} {1127}
  (\bibinfo {year} {2009})}\BibitemShut {NoStop}%
\bibitem [{\citenamefont {M\o{}gelh\o{}j}\ \emph {et~al.}(2011)\citenamefont
  {M\o{}gelh\o{}j}, \citenamefont {Kelkkanen}, \citenamefont {Wikfeldt},
  \citenamefont {Schi\o{}tz}, \citenamefont {Mortensen}, \citenamefont
  {Pettersson}, \citenamefont {Lundqvist}, \citenamefont {Jacobsen},
  \citenamefont {Nilsson},\ and\ \citenamefont {N\o{}rskov}}]{Mogelhoj2011}%
  \BibitemOpen
  \bibfield  {author} {\bibinfo {author} {\bibfnamefont {A.}~\bibnamefont
  {M\o{}gelh\o{}j}}, \bibinfo {author} {\bibfnamefont {A.~K.}\ \bibnamefont
  {Kelkkanen}}, \bibinfo {author} {\bibfnamefont {K.~T.}\ \bibnamefont
  {Wikfeldt}}, \bibinfo {author} {\bibfnamefont {J.}~\bibnamefont
  {Schi\o{}tz}}, \bibinfo {author} {\bibfnamefont {J.~J.}\ \bibnamefont
  {Mortensen}}, \bibinfo {author} {\bibfnamefont {L.~G.~M.}\ \bibnamefont
  {Pettersson}}, \bibinfo {author} {\bibfnamefont {B.~I.}\ \bibnamefont
  {Lundqvist}}, \bibinfo {author} {\bibfnamefont {K.~W.}\ \bibnamefont
  {Jacobsen}}, \bibinfo {author} {\bibfnamefont {A.}~\bibnamefont {Nilsson}}, \
  and\ \bibinfo {author} {\bibfnamefont {J.~K.}\ \bibnamefont {N\o{}rskov}},\
  }\href {\doibase 10.1021/jp2040345} {\bibfield  {journal} {\bibinfo
  {journal} {J. Phys. Chem. B}\ }\textbf {\bibinfo {volume} {115}},\ \bibinfo
  {pages} {14149} (\bibinfo {year} {2011})}\BibitemShut {NoStop}%
\bibitem [{\citenamefont {Wang}\ \emph {et~al.}(2011)\citenamefont {Wang},
  \citenamefont {Rom\'{a}n-P\'{e}rez}, \citenamefont {Soler}, \citenamefont
  {Artacho},\ and\ \citenamefont {Fern\'{a}ndez-Serra}}]{water_emiliomarivi}%
  \BibitemOpen
  \bibfield  {author} {\bibinfo {author} {\bibfnamefont {J.}~\bibnamefont
  {Wang}}, \bibinfo {author} {\bibfnamefont {G.}~\bibnamefont
  {Rom\'{a}n-P\'{e}rez}}, \bibinfo {author} {\bibfnamefont {J.~M.}\
  \bibnamefont {Soler}}, \bibinfo {author} {\bibfnamefont {E.}~\bibnamefont
  {Artacho}}, \ and\ \bibinfo {author} {\bibfnamefont {M.-V.}\ \bibnamefont
  {Fern\'{a}ndez-Serra}},\ }\href@noop {} {\bibfield  {journal} {\bibinfo
  {journal} {J. Chem. Phys.}\ }\textbf {\bibinfo {volume} {134}},\ \bibinfo
  {pages} {024516} (\bibinfo {year} {2011})}\BibitemShut {NoStop}%
\bibitem [{\citenamefont {Zhang}\ \emph {et~al.}(2011)\citenamefont {Zhang},
  \citenamefont {Wu}, \citenamefont {Galli},\ and\ \citenamefont
  {Gygi}}]{Zhang2011a}%
  \BibitemOpen
  \bibfield  {author} {\bibinfo {author} {\bibfnamefont {C.}~\bibnamefont
  {Zhang}}, \bibinfo {author} {\bibfnamefont {J.}~\bibnamefont {Wu}}, \bibinfo
  {author} {\bibfnamefont {G.}~\bibnamefont {Galli}}, \ and\ \bibinfo {author}
  {\bibfnamefont {F.}~\bibnamefont {Gygi}},\ }\href@noop {} {\bibfield
  {journal} {\bibinfo  {journal} {J. Chem. Theory Comput.}\ }\textbf {\bibinfo
  {volume} {7}},\ \bibinfo {pages} {3054} (\bibinfo {year} {2011})}\BibitemShut
  {NoStop}%
\bibitem [{\citenamefont {Wikfeldt}, \citenamefont {Nilsson},\ and\
  \citenamefont {Pettersson}(2011)}]{Wikfeldt2011}%
  \BibitemOpen
  \bibfield  {author} {\bibinfo {author} {\bibfnamefont {K.~T.}\ \bibnamefont
  {Wikfeldt}}, \bibinfo {author} {\bibfnamefont {A.}~\bibnamefont {Nilsson}}, \
  and\ \bibinfo {author} {\bibfnamefont {L.~G.~M.}\ \bibnamefont
  {Pettersson}},\ }\href@noop {} {\bibfield  {journal} {\bibinfo  {journal}
  {Phys. Chem. Chem. Phys.}\ }\textbf {\bibinfo {volume} {13}},\ \bibinfo
  {pages} {19918} (\bibinfo {year} {2011})}\BibitemShut {NoStop}%
\bibitem [{\citenamefont {Murray}\ and\ \citenamefont
  {Galli}(2012)}]{Murray2012}%
  \BibitemOpen
  \bibfield  {author} {\bibinfo {author} {\bibfnamefont {E.~D.}\ \bibnamefont
  {Murray}}\ and\ \bibinfo {author} {\bibfnamefont {G.}~\bibnamefont {Galli}},\
  }\href@noop {} {\bibfield  {journal} {\bibinfo  {journal} {Phys. Rev. Lett.}\
  }\textbf {\bibinfo {volume} {108}},\ \bibinfo {pages} {105502} (\bibinfo
  {year} {2012})}\BibitemShut {NoStop}%
\bibitem [{\citenamefont {Soler}\ \emph {et~al.}(2002)\citenamefont {Soler},
  \citenamefont {Artacho}, \citenamefont {Gale}, \citenamefont {Garc\'{\i}a},
  \citenamefont {Junquera}, \citenamefont {Ordej\'{o}n},\ and\ \citenamefont
  {S\'{a}nchez-Portal}}]{Soler2002}%
  \BibitemOpen
  \bibfield  {author} {\bibinfo {author} {\bibfnamefont {J.~M.}\ \bibnamefont
  {Soler}}, \bibinfo {author} {\bibfnamefont {E.}~\bibnamefont {Artacho}},
  \bibinfo {author} {\bibfnamefont {J.~D.}\ \bibnamefont {Gale}}, \bibinfo
  {author} {\bibfnamefont {A.}~\bibnamefont {Garc\'{\i}a}}, \bibinfo {author}
  {\bibfnamefont {J.}~\bibnamefont {Junquera}}, \bibinfo {author}
  {\bibfnamefont {P.}~\bibnamefont {Ordej\'{o}n}}, \ and\ \bibinfo {author}
  {\bibfnamefont {D.}~\bibnamefont {S\'{a}nchez-Portal}},\ }\href@noop {}
  {\bibfield  {journal} {\bibinfo  {journal} {J. Phys.: Condens. Matter}\
  }\textbf {\bibinfo {volume} {14}},\ \bibinfo {pages} {2745} (\bibinfo {year}
  {2002})}\BibitemShut {NoStop}%
\bibitem [{\citenamefont {Troullier}\ and\ \citenamefont
  {Martins}(1991)}]{Troullier1991}%
  \BibitemOpen
  \bibfield  {author} {\bibinfo {author} {\bibfnamefont {N.}~\bibnamefont
  {Troullier}}\ and\ \bibinfo {author} {\bibfnamefont {J.~L.}\ \bibnamefont
  {Martins}},\ }\href@noop {} {\bibfield  {journal} {\bibinfo  {journal} {Phys.
  Rev. B}\ }\textbf {\bibinfo {volume} {43}},\ \bibinfo {pages} {1993}
  (\bibinfo {year} {1991})}\BibitemShut {NoStop}%
\bibitem [{\citenamefont {Junquera}\ \emph {et~al.}(2001)\citenamefont
  {Junquera}, \citenamefont {Paz}, \citenamefont {S\'{a}nchez-Portal},\ and\
  \citenamefont {Artacho}}]{Junquera2001}%
  \BibitemOpen
  \bibfield  {author} {\bibinfo {author} {\bibfnamefont {J.}~\bibnamefont
  {Junquera}}, \bibinfo {author} {\bibfnamefont {O.}~\bibnamefont {Paz}},
  \bibinfo {author} {\bibfnamefont {D.}~\bibnamefont {S\'{a}nchez-Portal}}, \
  and\ \bibinfo {author} {\bibfnamefont {E.}~\bibnamefont {Artacho}},\
  }\href@noop {} {\bibfield  {journal} {\bibinfo  {journal} {Phys. Rev. B}\
  }\textbf {\bibinfo {volume} {64}},\ \bibinfo {pages} {235111} (\bibinfo
  {year} {2001})}\BibitemShut {NoStop}%
\bibitem [{\citenamefont {Anglada}\ \emph {et~al.}(2002)\citenamefont
  {Anglada}, \citenamefont {Soler}, \citenamefont {Junquera},\ and\
  \citenamefont {Artacho}}]{Anglada2002}%
  \BibitemOpen
  \bibfield  {author} {\bibinfo {author} {\bibfnamefont {E.}~\bibnamefont
  {Anglada}}, \bibinfo {author} {\bibfnamefont {J.~M.}\ \bibnamefont {Soler}},
  \bibinfo {author} {\bibfnamefont {J.}~\bibnamefont {Junquera}}, \ and\
  \bibinfo {author} {\bibfnamefont {E.}~\bibnamefont {Artacho}},\ }\href@noop
  {} {\bibfield  {journal} {\bibinfo  {journal} {Phys. Rev. B}\ }\textbf
  {\bibinfo {volume} {66}},\ \bibinfo {pages} {205101} (\bibinfo {year}
  {2002})}\BibitemShut {NoStop}%
\bibitem [{\citenamefont {Corsetti}\ \emph {et~al.}(2013)\citenamefont
  {Corsetti}, \citenamefont {Fern\'{a}ndez-Serra}, \citenamefont {Soler},\ and\
  \citenamefont {Artacho}}]{Corsetti2013}%
  \BibitemOpen
  \bibfield  {author} {\bibinfo {author} {\bibfnamefont {F.}~\bibnamefont
  {Corsetti}}, \bibinfo {author} {\bibfnamefont {M.-V.}\ \bibnamefont
  {Fern\'{a}ndez-Serra}}, \bibinfo {author} {\bibfnamefont {J.~M.}\
  \bibnamefont {Soler}}, \ and\ \bibinfo {author} {\bibfnamefont
  {E.}~\bibnamefont {Artacho}},\ }\href@noop {} {\bibfield  {journal} {\bibinfo
   {journal} {J. Phys.: Condens. Matter}\ }\textbf {\bibinfo {volume} {25}},\
  \bibinfo {pages} {435504} (\bibinfo {year} {2013})}\BibitemShut {NoStop}%
\bibitem [{\citenamefont {Jorgensen}\ \emph {et~al.}(1983)\citenamefont
  {Jorgensen}, \citenamefont {Chandrasekhar}, \citenamefont {Madura},
  \citenamefont {Impey},\ and\ \citenamefont {Klein}}]{Jorgensen1983b}%
  \BibitemOpen
  \bibfield  {author} {\bibinfo {author} {\bibfnamefont {W.~L.}\ \bibnamefont
  {Jorgensen}}, \bibinfo {author} {\bibfnamefont {J.}~\bibnamefont
  {Chandrasekhar}}, \bibinfo {author} {\bibfnamefont {J.~D.}\ \bibnamefont
  {Madura}}, \bibinfo {author} {\bibfnamefont {R.~W.}\ \bibnamefont {Impey}}, \
  and\ \bibinfo {author} {\bibfnamefont {M.~L.}\ \bibnamefont {Klein}},\
  }\href@noop {} {\bibfield  {journal} {\bibinfo  {journal} {J. Chem. Phys.}\
  }\textbf {\bibinfo {volume} {79}},\ \bibinfo {pages} {926} (\bibinfo {year}
  {1983})}\BibitemShut {NoStop}%
\bibitem [{\citenamefont {Berendsen}, \citenamefont {van~der Spoel},\ and\
  \citenamefont {van Drunen}(1995)}]{gromacs}%
  \BibitemOpen
  \bibfield  {author} {\bibinfo {author} {\bibfnamefont {H.~J.~C.}\
  \bibnamefont {Berendsen}}, \bibinfo {author} {\bibfnamefont {D.}~\bibnamefont
  {van~der Spoel}}, \ and\ \bibinfo {author} {\bibfnamefont {R.}~\bibnamefont
  {van Drunen}},\ }\href@noop {} {\bibfield  {journal} {\bibinfo  {journal}
  {Comput. Phys. Commun.}\ }\textbf {\bibinfo {volume} {91}},\ \bibinfo {pages}
  {43} (\bibinfo {year} {1995})}\BibitemShut {NoStop}%
\bibitem [{\citenamefont {K\"{u}hne}, \citenamefont {Krack},\ and\
  \citenamefont {Parrinello}(2009)}]{Kuhne2009}%
  \BibitemOpen
  \bibfield  {author} {\bibinfo {author} {\bibfnamefont {T.~D.}\ \bibnamefont
  {K\"{u}hne}}, \bibinfo {author} {\bibfnamefont {M.}~\bibnamefont {Krack}}, \
  and\ \bibinfo {author} {\bibfnamefont {M.}~\bibnamefont {Parrinello}},\
  }\href@noop {} {\bibfield  {journal} {\bibinfo  {journal} {J. Chem. Theory
  Comput.}\ }\textbf {\bibinfo {volume} {5}},\ \bibinfo {pages} {235} (\bibinfo
  {year} {2009})}\BibitemShut {NoStop}%
\bibitem [{\citenamefont {Gonze~{\em et al.}}(2009)}]{abinit-generic-et-al2}%
  \BibitemOpen
  \bibfield  {author} {\bibinfo {author} {\bibfnamefont {X.}~\bibnamefont
  {Gonze~{\em et al.}}},\ }\href@noop {} {\bibfield  {journal} {\bibinfo
  {journal} {Comput. Phys. Commun.}\ }\textbf {\bibinfo {volume} {180}},\
  \bibinfo {pages} {2582} (\bibinfo {year} {2009})}\BibitemShut {NoStop}%
\bibitem [{\citenamefont {Kleinman}\ and\ \citenamefont
  {Bylander}(1982)}]{pseudo-KB}%
  \BibitemOpen
  \bibfield  {author} {\bibinfo {author} {\bibfnamefont {L.}~\bibnamefont
  {Kleinman}}\ and\ \bibinfo {author} {\bibfnamefont {D.~M.}\ \bibnamefont
  {Bylander}},\ }\href {\doibase 10.1103/PhysRevLett.48.1425} {\bibfield
  {journal} {\bibinfo  {journal} {Phys. Rev. Lett.}\ }\textbf {\bibinfo
  {volume} {48}},\ \bibinfo {pages} {1425} (\bibinfo {year}
  {1982})}\BibitemShut {NoStop}%
\bibitem [{\citenamefont {Meade}\ and\ \citenamefont
  {Vanderbilt}(1989)}]{Meade1989}%
  \BibitemOpen
  \bibfield  {author} {\bibinfo {author} {\bibfnamefont {R.~D.}\ \bibnamefont
  {Meade}}\ and\ \bibinfo {author} {\bibfnamefont {D.}~\bibnamefont
  {Vanderbilt}},\ }\href@noop {} {\bibfield  {journal} {\bibinfo  {journal}
  {Mat. Res. Soc. Symp. Proc.}\ }\textbf {\bibinfo {volume} {141}},\ \bibinfo
  {pages} {451} (\bibinfo {year} {1989})}\BibitemShut {NoStop}%
\bibitem [{\citenamefont {Zhang}\ and\ \citenamefont {Yang}(1998)}]{revPBE}%
  \BibitemOpen
  \bibfield  {author} {\bibinfo {author} {\bibfnamefont {Y.}~\bibnamefont
  {Zhang}}\ and\ \bibinfo {author} {\bibfnamefont {W.}~\bibnamefont {Yang}},\
  }\href@noop {} {\bibfield  {journal} {\bibinfo  {journal} {J. Chem. Phys.}\
  }\textbf {\bibinfo {volume} {80}},\ \bibinfo {pages} {890} (\bibinfo {year}
  {1998})}\BibitemShut {NoStop}%
\bibitem [{\citenamefont {Perdew}, \citenamefont {Burke},\ and\ \citenamefont
  {Ernzerhof}(1996)}]{pbe}%
  \BibitemOpen
  \bibfield  {author} {\bibinfo {author} {\bibfnamefont {J.~P.}\ \bibnamefont
  {Perdew}}, \bibinfo {author} {\bibfnamefont {K.}~\bibnamefont {Burke}}, \
  and\ \bibinfo {author} {\bibfnamefont {M.}~\bibnamefont {Ernzerhof}},\
  }\href@noop {} {\bibfield  {journal} {\bibinfo  {journal} {Phys. Rev. Lett.}\
  }\textbf {\bibinfo {volume} {77}},\ \bibinfo {pages} {3865} (\bibinfo {year}
  {1996})}\BibitemShut {NoStop}%
\bibitem [{\citenamefont {Murray}, \citenamefont {Lee},\ and\ \citenamefont
  {Langreth}(2009)}]{PW86RPBE}%
  \BibitemOpen
  \bibfield  {author} {\bibinfo {author} {\bibfnamefont {{\'{E}}.~D.}\
  \bibnamefont {Murray}}, \bibinfo {author} {\bibfnamefont {K.}~\bibnamefont
  {Lee}}, \ and\ \bibinfo {author} {\bibfnamefont {D.~C.}\ \bibnamefont
  {Langreth}},\ }\href@noop {} {\bibfield  {journal} {\bibinfo  {journal} {J.
  Chem. Theory Comput.}\ }\textbf {\bibinfo {volume} {5}},\ \bibinfo {pages}
  {2754} (\bibinfo {year} {2009})}\BibitemShut {NoStop}%
\bibitem [{\citenamefont {Rom\'{a}n-P\'{e}rez}\ and\ \citenamefont
  {Soler}(2009)}]{RomanPerez2009}%
  \BibitemOpen
  \bibfield  {author} {\bibinfo {author} {\bibfnamefont {G.}~\bibnamefont
  {Rom\'{a}n-P\'{e}rez}}\ and\ \bibinfo {author} {\bibfnamefont {J.~M.}\
  \bibnamefont {Soler}},\ }\href@noop {} {\bibfield  {journal} {\bibinfo
  {journal} {Phys. Rev. Lett.}\ }\textbf {\bibinfo {volume} {103}},\ \bibinfo
  {pages} {096102} (\bibinfo {year} {2009})}\BibitemShut {NoStop}%
\bibitem [{\citenamefont {Sabatini}, \citenamefont {Gorni},\ and\ \citenamefont
  {de~Gironcoli}(2013)}]{Sabatini2013}%
  \BibitemOpen
  \bibfield  {author} {\bibinfo {author} {\bibfnamefont {R.}~\bibnamefont
  {Sabatini}}, \bibinfo {author} {\bibfnamefont {T.}~\bibnamefont {Gorni}}, \
  and\ \bibinfo {author} {\bibfnamefont {S.}~\bibnamefont {de~Gironcoli}},\
  }\href@noop {} {\bibfield  {journal} {\bibinfo  {journal} {Phys. Rev. B}\
  }\textbf {\bibinfo {volume} {87}},\ \bibinfo {pages} {041108(R)} (\bibinfo
  {year} {2013})}\BibitemShut {NoStop}%
\bibitem [{\citenamefont {Wu}\ and\ \citenamefont {Gygi}(2012)}]{Wu-Gygi2012}%
  \BibitemOpen
  \bibfield  {author} {\bibinfo {author} {\bibfnamefont {J.}~\bibnamefont
  {Wu}}\ and\ \bibinfo {author} {\bibfnamefont {F.}~\bibnamefont {Gygi}},\
  }\href@noop {} {\bibfield  {journal} {\bibinfo  {journal} {J. Chem. Phys.}\
  }\textbf {\bibinfo {volume} {136}},\ \bibinfo {pages} {224107} (\bibinfo
  {year} {2012})}\BibitemShut {NoStop}%
\bibitem [{\citenamefont {Wagner}\ and\ \citenamefont
  {Pru{\ss}}(2002)}]{water_density}%
  \BibitemOpen
  \bibfield  {author} {\bibinfo {author} {\bibfnamefont {W.}~\bibnamefont
  {Wagner}}\ and\ \bibinfo {author} {\bibfnamefont {A.}~\bibnamefont
  {Pru{\ss}}},\ }\href@noop {} {\bibfield  {journal} {\bibinfo  {journal} {J.
  Phys. Chem. Ref. Data}\ }\textbf {\bibinfo {volume} {31}},\ \bibinfo {pages}
  {387} (\bibinfo {year} {2002})}\BibitemShut {NoStop}%
\bibitem [{\citenamefont {Mahoney}\ and\ \citenamefont
  {Jorgensen}(2000)}]{TIP5P}%
  \BibitemOpen
  \bibfield  {author} {\bibinfo {author} {\bibfnamefont {M.~W.}\ \bibnamefont
  {Mahoney}}\ and\ \bibinfo {author} {\bibfnamefont {W.~L.}\ \bibnamefont
  {Jorgensen}},\ }\href@noop {} {\bibfield  {journal} {\bibinfo  {journal} {J.
  Chem. Phys.}\ }\textbf {\bibinfo {volume} {112}},\ \bibinfo {pages} {8910}
  (\bibinfo {year} {2000})}\BibitemShut {NoStop}%
\bibitem [{\citenamefont {Berendsen}, \citenamefont {Grigera},\ and\
  \citenamefont {Straatsma}(1987)}]{SPCE}%
  \BibitemOpen
  \bibfield  {author} {\bibinfo {author} {\bibfnamefont {H.~J.~C.}\
  \bibnamefont {Berendsen}}, \bibinfo {author} {\bibfnamefont {J.~R.}\
  \bibnamefont {Grigera}}, \ and\ \bibinfo {author} {\bibfnamefont {T.~P.}\
  \bibnamefont {Straatsma}},\ }\href@noop {} {\bibfield  {journal} {\bibinfo
  {journal} {J. Phys. Chem.}\ }\textbf {\bibinfo {volume} {91}},\ \bibinfo
  {pages} {6269} (\bibinfo {year} {1987})}\BibitemShut {NoStop}%
\bibitem [{\citenamefont {Skinner}\ \emph {et~al.}(2013)\citenamefont
  {Skinner}, \citenamefont {Huang}, \citenamefont {Schlesinger}, \citenamefont
  {Pettersson}, \citenamefont {Nilsson},\ and\ \citenamefont
  {Benmore}}]{waterRDF2013}%
  \BibitemOpen
  \bibfield  {author} {\bibinfo {author} {\bibfnamefont {L.~B.}\ \bibnamefont
  {Skinner}}, \bibinfo {author} {\bibfnamefont {C.}~\bibnamefont {Huang}},
  \bibinfo {author} {\bibfnamefont {D.}~\bibnamefont {Schlesinger}}, \bibinfo
  {author} {\bibfnamefont {L.~G.~M.}\ \bibnamefont {Pettersson}}, \bibinfo
  {author} {\bibfnamefont {A.}~\bibnamefont {Nilsson}}, \ and\ \bibinfo
  {author} {\bibfnamefont {C.~J.}\ \bibnamefont {Benmore}},\ }\href@noop {}
  {\bibfield  {journal} {\bibinfo  {journal} {J. Chem. Phys.}\ }\textbf
  {\bibinfo {volume} {138}},\ \bibinfo {pages} {074506} (\bibinfo {year}
  {2013})}\BibitemShut {NoStop}%
\bibitem [{\citenamefont {Soper}\ and\ \citenamefont {Ricci}(2000)}]{soper}%
  \BibitemOpen
  \bibfield  {author} {\bibinfo {author} {\bibfnamefont {A.~K.}\ \bibnamefont
  {Soper}}\ and\ \bibinfo {author} {\bibfnamefont {M.~A.}\ \bibnamefont
  {Ricci}},\ }\href@noop {} {\bibfield  {journal} {\bibinfo  {journal} {Phys.
  Rev. Lett.}\ }\textbf {\bibinfo {volume} {84}},\ \bibinfo {pages} {2881}
  (\bibinfo {year} {2000})}\BibitemShut {NoStop}%
\bibitem [{Note1()}]{Note1}%
  \BibitemOpen
  \bibinfo {note} {Interestingly, a recent classical force-field
  simulation~\cite {Kaya2013} of thin-film water on a BaF$_2$ surface has
  reported a very similar extended minimum in the RDF for the first 1~\r
  A-thick water layer above the surface, which is therefore suggested to be in
  a highly compressed state similar to HDL.}\BibitemShut {Stop}%
\bibitem [{Note2()}]{Note2}%
  \BibitemOpen
  \bibinfo {note} {The $P$--$\rho $ curves in Fig.~\ref {fig:EoS} arguably
  reveal the same behavior: by extrapolating the vdW-DF$^\protect \mathrm
  {PBE}$ curve, we can see a tentative crossing with experiment around
  1.25~g/cm$^3$ and, analogously, a crossing of VV10 with experiment at low
  density (the precise value being harder to estimate in this
  case).}\BibitemShut {Stop}%
\bibitem [{\citenamefont {Krynicki}, \citenamefont {Green},\ and\ \citenamefont
  {Sawyer}(1978)}]{H20_diff_max}%
  \BibitemOpen
  \bibfield  {author} {\bibinfo {author} {\bibfnamefont {K.}~\bibnamefont
  {Krynicki}}, \bibinfo {author} {\bibfnamefont {C.~D.}\ \bibnamefont {Green}},
  \ and\ \bibinfo {author} {\bibfnamefont {D.~W.}\ \bibnamefont {Sawyer}},\
  }\href@noop {} {\bibfield  {journal} {\bibinfo  {journal} {Faraday Discuss.
  Chem. Soc.}\ }\textbf {\bibinfo {volume} {66}},\ \bibinfo {pages} {199}
  (\bibinfo {year} {1978})}\BibitemShut {NoStop}%
\bibitem [{\citenamefont {Mills}(1973)}]{H2O_D2O_diff}%
  \BibitemOpen
  \bibfield  {author} {\bibinfo {author} {\bibfnamefont {R.}~\bibnamefont
  {Mills}},\ }\href@noop {} {\bibfield  {journal} {\bibinfo  {journal} {J.
  Phys. Chem.}\ }\textbf {\bibinfo {volume} {77}},\ \bibinfo {pages} {685}
  (\bibinfo {year} {1973})}\BibitemShut {NoStop}%
\bibitem [{\citenamefont {Scala}\ \emph {et~al.}(2000)\citenamefont {Scala},
  \citenamefont {Starr}, \citenamefont {{La Nave}}, \citenamefont {Sciortino},\
  and\ \citenamefont {Stanley}}]{Scala2000}%
  \BibitemOpen
  \bibfield  {author} {\bibinfo {author} {\bibfnamefont {A.}~\bibnamefont
  {Scala}}, \bibinfo {author} {\bibfnamefont {F.~W.}\ \bibnamefont {Starr}},
  \bibinfo {author} {\bibfnamefont {E.}~\bibnamefont {{La Nave}}}, \bibinfo
  {author} {\bibfnamefont {F.}~\bibnamefont {Sciortino}}, \ and\ \bibinfo
  {author} {\bibfnamefont {H.~E.}\ \bibnamefont {Stanley}},\ }\href@noop {}
  {\bibfield  {journal} {\bibinfo  {journal} {Nature}\ }\textbf {\bibinfo
  {volume} {406}},\ \bibinfo {pages} {166} (\bibinfo {year}
  {2000})}\BibitemShut {NoStop}%
\bibitem [{\citenamefont {Errington}\ and\ \citenamefont
  {Debenedetti}(2001)}]{Errington2001}%
  \BibitemOpen
  \bibfield  {author} {\bibinfo {author} {\bibfnamefont {J.~R.}\ \bibnamefont
  {Errington}}\ and\ \bibinfo {author} {\bibfnamefont {P.~G.}\ \bibnamefont
  {Debenedetti}},\ }\href@noop {} {\bibfield  {journal} {\bibinfo  {journal}
  {Nature}\ }\textbf {\bibinfo {volume} {409}},\ \bibinfo {pages} {318}
  (\bibinfo {year} {2001})}\BibitemShut {NoStop}%
\bibitem [{\citenamefont {Netz}\ \emph {et~al.}(2002)\citenamefont {Netz},
  \citenamefont {Starr}, \citenamefont {Barbosa},\ and\ \citenamefont
  {Stanley}}]{Netz2002}%
  \BibitemOpen
  \bibfield  {author} {\bibinfo {author} {\bibfnamefont {P.~A.}\ \bibnamefont
  {Netz}}, \bibinfo {author} {\bibfnamefont {F.}~\bibnamefont {Starr}},
  \bibinfo {author} {\bibfnamefont {M.~C.}\ \bibnamefont {Barbosa}}, \ and\
  \bibinfo {author} {\bibfnamefont {H.~E.}\ \bibnamefont {Stanley}},\
  }\href@noop {} {\bibfield  {journal} {\bibinfo  {journal} {J. Mol. Liq.}\
  }\textbf {\bibinfo {volume} {101}},\ \bibinfo {pages} {159} (\bibinfo {year}
  {2002})}\BibitemShut {NoStop}%
\bibitem [{\citenamefont {D\"{u}nweg}\ and\ \citenamefont
  {Kremer}(1993)}]{Dunweg1993}%
  \BibitemOpen
  \bibfield  {author} {\bibinfo {author} {\bibfnamefont {B.}~\bibnamefont
  {D\"{u}nweg}}\ and\ \bibinfo {author} {\bibfnamefont {K.}~\bibnamefont
  {Kremer}},\ }\href@noop {} {\bibfield  {journal} {\bibinfo  {journal} {J.
  Chem. Phys.}\ }\textbf {\bibinfo {volume} {99}},\ \bibinfo {pages} {6983}
  (\bibinfo {year} {1993})}\BibitemShut {NoStop}%
\bibitem [{\citenamefont {Wilbur}, \citenamefont {DeFries},\ and\ \citenamefont
  {Jonas}(1976)}]{D20_diff_max}%
  \BibitemOpen
  \bibfield  {author} {\bibinfo {author} {\bibfnamefont {D.~J.}\ \bibnamefont
  {Wilbur}}, \bibinfo {author} {\bibfnamefont {T.}~\bibnamefont {DeFries}}, \
  and\ \bibinfo {author} {\bibfnamefont {J.}~\bibnamefont {Jonas}},\
  }\href@noop {} {\bibfield  {journal} {\bibinfo  {journal} {J. Chem. Phys.}\
  }\textbf {\bibinfo {volume} {65}},\ \bibinfo {pages} {1783} (\bibinfo {year}
  {1976})}\BibitemShut {NoStop}%
\bibitem [{\citenamefont {Perdew}\ and\ \citenamefont {Zunger}(1981)}]{qmc2}%
  \BibitemOpen
  \bibfield  {author} {\bibinfo {author} {\bibfnamefont {J.~P.}\ \bibnamefont
  {Perdew}}\ and\ \bibinfo {author} {\bibfnamefont {A.}~\bibnamefont
  {Zunger}},\ }\href@noop {} {\bibfield  {journal} {\bibinfo  {journal} {Phys.
  Rev. B}\ }\textbf {\bibinfo {volume} {23}},\ \bibinfo {pages} {5048}
  (\bibinfo {year} {1981})}\BibitemShut {NoStop}%
\bibitem [{\citenamefont {Fern\'andez-Serra}\ and\ \citenamefont
  {Artacho}(2006)}]{Fernandez-Serra2006}%
  \BibitemOpen
  \bibfield  {author} {\bibinfo {author} {\bibfnamefont {M.~V.}\ \bibnamefont
  {Fern\'andez-Serra}}\ and\ \bibinfo {author} {\bibfnamefont {E.}~\bibnamefont
  {Artacho}},\ }\href@noop {} {\bibfield  {journal} {\bibinfo  {journal} {Phys.
  Rev. Lett.}\ }\textbf {\bibinfo {volume} {96}},\ \bibinfo {pages} {016404}
  (\bibinfo {year} {2006})}\BibitemShut {NoStop}%
\bibitem [{\citenamefont {Kumar}, \citenamefont {Schmidt},\ and\ \citenamefont
  {Skinner}(2007)}]{Kumar2007}%
  \BibitemOpen
  \bibfield  {author} {\bibinfo {author} {\bibfnamefont {R.}~\bibnamefont
  {Kumar}}, \bibinfo {author} {\bibfnamefont {J.~R.}\ \bibnamefont {Schmidt}},
  \ and\ \bibinfo {author} {\bibfnamefont {J.~L.}\ \bibnamefont {Skinner}},\
  }\href@noop {} {\bibfield  {journal} {\bibinfo  {journal} {J. Chem. Phys.}\
  }\textbf {\bibinfo {volume} {126}},\ \bibinfo {pages} {204107} (\bibinfo
  {year} {2007})}\BibitemShut {NoStop}%
\bibitem [{\citenamefont {Sharma}, \citenamefont {Resta},\ and\ \citenamefont
  {Car}(2005)}]{Sharma2005}%
  \BibitemOpen
  \bibfield  {author} {\bibinfo {author} {\bibfnamefont {M.}~\bibnamefont
  {Sharma}}, \bibinfo {author} {\bibfnamefont {R.}~\bibnamefont {Resta}}, \
  and\ \bibinfo {author} {\bibfnamefont {R.}~\bibnamefont {Car}},\ }\href@noop
  {} {\bibfield  {journal} {\bibinfo  {journal} {Phys. Rev. Lett.}\ }\textbf
  {\bibinfo {volume} {95}},\ \bibinfo {pages} {187401} (\bibinfo {year}
  {2005})}\BibitemShut {NoStop}%
\bibitem [{\citenamefont {Stillinger}\ and\ \citenamefont
  {Rahman}(1974)}]{ST2}%
  \BibitemOpen
  \bibfield  {author} {\bibinfo {author} {\bibfnamefont {F.~H.}\ \bibnamefont
  {Stillinger}}\ and\ \bibinfo {author} {\bibfnamefont {A.}~\bibnamefont
  {Rahman}},\ }\href@noop {} {\bibfield  {journal} {\bibinfo  {journal} {J.
  Chem. Phys.}\ }\textbf {\bibinfo {volume} {60}},\ \bibinfo {pages} {1545}
  (\bibinfo {year} {1974})}\BibitemShut {NoStop}%
\bibitem [{\citenamefont {Kaya}\ \emph {et~al.}(2013)\citenamefont {Kaya},
  \citenamefont {Schlesinger}, \citenamefont {Yamamoto}, \citenamefont
  {Newberg}, \citenamefont {Bluhm}, \citenamefont {Ogasawara}, \citenamefont
  {Kendelewicz}, \citenamefont {{Brown Jr.}}, \citenamefont {Pettersson},\ and\
  \citenamefont {Nilsson}}]{Kaya2013}%
  \BibitemOpen
  \bibfield  {author} {\bibinfo {author} {\bibfnamefont {S.}~\bibnamefont
  {Kaya}}, \bibinfo {author} {\bibfnamefont {D.}~\bibnamefont {Schlesinger}},
  \bibinfo {author} {\bibfnamefont {S.}~\bibnamefont {Yamamoto}}, \bibinfo
  {author} {\bibfnamefont {J.~T.}\ \bibnamefont {Newberg}}, \bibinfo {author}
  {\bibfnamefont {H.}~\bibnamefont {Bluhm}}, \bibinfo {author} {\bibfnamefont
  {H.}~\bibnamefont {Ogasawara}}, \bibinfo {author} {\bibfnamefont
  {T.}~\bibnamefont {Kendelewicz}}, \bibinfo {author} {\bibfnamefont {G.~E.}\
  \bibnamefont {{Brown Jr.}}}, \bibinfo {author} {\bibfnamefont {L.~G.~M.}\
  \bibnamefont {Pettersson}}, \ and\ \bibinfo {author} {\bibfnamefont
  {A.}~\bibnamefont {Nilsson}},\ }\href@noop {} {\bibfield  {journal} {\bibinfo
   {journal} {Sci. Rep.}\ }\textbf {\bibinfo {volume} {3}},\ \bibinfo {pages}
  {1074} (\bibinfo {year} {2013})}\BibitemShut {NoStop}%
\end{thebibliography}
\end{document}